\documentclass[journal]{IEEEtran}
\usepackage{amsfonts}
\usepackage{bbm}
\usepackage{lettrine}
\usepackage{pdfcomment}

\usepackage{ifpdf}

\usepackage{xcolor}
\usepackage{soul}
\usepackage{cite}
\usepackage{balance}
\usepackage{hyperref}

\hypersetup{colorlinks,linkcolor={black},citecolor={black},urlcolor={black}}

\ifCLASSINFOpdf
   \usepackage[pdftex]{graphicx}
\else
   \usepackage[dvips]{graphicx}
\fi
\usepackage{color}
\usepackage{amsmath}
\usepackage{amsthm}
\usepackage{bm}
\newtheorem{thm}{Theorem}

\DeclareMathOperator*{\argmax}{arg\,max}

\usepackage[linesnumbered,ruled,vlined]{algorithm2e}
\let\oldnl\nl
\newcommand{\nonl}{\renewcommand{\nl}{\let\nl\oldnl}}
\SetKwInput{KwInput}{Input}
\SetKwInput{KwOutput}{Output}

\usepackage{algorithmicx, algpseudocode}
\usepackage{diagbox}
\usepackage{array}

\usepackage{array}
\usepackage{fixltx2e}

\usepackage{stfloats}
\usepackage{url}

\usepackage{pifont}
\usepackage{makecell}

\begin{document}

\title{\Huge{Open RAN-Based Mixed-Timescale and Robust Task Offloading in Vehicular Edge Computing}}
%
%
%

\author{Mohsen~Tajallifar,~Nizar~Zorba,~\IEEEmembership{Senior Member,~IEEE}, Nader~Mokari,~\IEEEmembership{Senior Member,~IEEE,~and~Hamid~Saeedi,~\IEEEmembership{Member,~IEEE}}
\thanks{M. Tajallifar and N. Zorba
are with College of Engineering, Qatar University, Doha, Qatar,
(e-mail: \{mohsen.tajallifar,nizarz\}@qu.edu.qr).
N. Mokari is with ECE Department, Tarbiat Modares University, Tehran, Iran (e-mail: nader.mokari@modares.ac.ir). H. Saeedi is with College of Engineering and Technology, University of Doha for Science and Technology, Doha, Qatar, (e-mail: hsaeedi@udst.edu.qa).}
\thanks{A preliminary version of this work was presented at the IEEE Global Communications Conference (GLOBECOM), Taiwan, December 2025.}
\thanks{This work was supported by the Qatar Research, Development and Innovation (QRDI) Council under Grant ARG01-0511-230129. }
}
\maketitle
\begin{abstract}
Vehicular edge computing (VEC) emerges as a critical enabler of low-latency and computation-intensive vehicular applications by offloading tasks from vehicles to edge servers. However, the dynamic nature of vehicular networks introduces significant uncertainty in task characteristics and network conditions, posing challenges for efficient resource allocation. This paper proposes a robust task offloading scheme for VEC within the open radio access network (O-RAN) architecture. Leveraging mixed-timescale optimization, the proposed approach integrates large-timescale computational resource allocation (CRA), moderate-timescale task partitioning, and small-timescale radio resource allocation (RRA) using O-RAN's hierarchical control framework. The robust optimization model minimizes the network-wide resources while meeting the deadline of individual task execution latencies under demand uncertainty. We employ the cutting-set method to address the demand uncertainty in the large-timescale CRA. We obtain a closed-form solution to the optimal task partitioning problem. We adopt both the Lyapunov optimization and a heuristic approach for the small-timescale RRA, wherein we adopt the concave-convex procedure to resolve the non-convexity of the data rate expression. The efficiency of the proposed method is shown through extensive simulations. The results show that the small-timescale RRA succeeds to counteract the demand uncertainty at the large-timescale CRA\textcolor{black}{, that is, no outage occurs when demands are in the assumed uncertainty set, whereas the non-robust scheme exhibits as high as $50 \%$ outage probability}. Moreover, our heuristic scheme is less conservative and consumes less resources than the Lyapunov optimization\textcolor{black}{, and it consumes about $40 \%$ less bandwidth than the conventional non-slotted robust solution.}
\end{abstract}


\begin{IEEEkeywords}
Vehicular edge computing, task offloading, demand uncertainty, mixed-timescale optimization, robust optimization
\end{IEEEkeywords}

%
\IEEEpeerreviewmaketitle
{\color{black}
\section{Introduction}\label{introduction_sec}
\lettrine{V}{ehicular} networks are a crucial enabler of intelligent transportation systems, enhancing road safety and traffic efficiency. These networks support services such as safety-critical, traffic efficiency, and infotainment by exchanging information between vehicles, roadside infrastructure, pedestrians, etc.\ \cite{clancy2024wireless, parvini2023aoi}. Many vehicular applications are both computation-intensive and delay-sensitive. Applications such as autonomous driving, real-time video analytics, and augmented reality require significant computational resources to process large data volumes and demand ultra-low latency to ensure timely responses, as even slight delays can lead to catastrophic consequences in safety-critical scenarios. Meeting these requirements is challenging due to vehicles’ limited on-board computational capabilities \cite{duan2020emerging}.

Vehicular edge computing (VEC) has emerged as a promising solution to address these computational and latency challenges \cite{liu2021vehicular, zhou2020evolutionary}. VEC leverages edge servers at the network’s edge, such as roadside units (RSUs) and base stations (BSs), to provide low-latency, computation-intensive services. By offloading tasks to nearby edge servers, VEC reduces latency and enhances the available computational capacity, making it suitable for many vehicular applications.

Task offloading is a key technique for realizing VEC's benefits. Vehicles can execute tasks locally or offload them to VEC servers depending on factors such as computational requirements and network conditions.

The performance of vehicular applications depends on experienced latency, and standards define a maximum latency for each application \cite{3gppV2X, 3gppenhanced}. Thus, task completion must occur within the specified value. However, conventional task offloading schemes minimizing latency or joint latency-energy measures may not guarantee completion within acceptable latency. Energy constraints are also less stringent than in cellular devices. Instead, service providers aim to minimize network-wide resource consumption, such as air interface bandwidth and VEC server computational capacity, while ensuring timely task completion.

The effectiveness of task offloading depends heavily on task characteristics such as input data size, required computational resources, and the maximum acceptable latency, i.e., the task deadline. Conventional schemes typically assume that task characteristics are fully known during decision-making, which is often unrealistic in dynamic vehicular environments. When a vehicle enters the coverage of an edge server and requests a vehicular application comprising successive tasks with sub-second intervals, these tasks are periodically generated and may change over time due to the environment’s high dynamism.

To serve such applications, computational resources at the edge server are generally allocated long-term, i.e., cannot be updated in millisecond range because virtual machine (VM) setup and reconfiguration are time-consuming \cite{xia2018power, tang2019systematic}. As a result, computational resource allocation (CRA) is performed in a large-timescale, e.g., for the entire sojourn time of vehicles. In this scenario, tasks' characteristics cannot be known at the time of CRA due to sub-second task renewal, necessitating robust task offloading schemes that handle variations without violating latency deadlines.

To mitigate uncertainty in task offloading, robust optimization is often employed \cite{li2020data, nguyen2021two}, aiming to make decisions robust to worst-case scenarios. However, robust optimization tends to be conservative, over-provisioning computational resources to avoid deadline violations. While ensuring reliability, this can cause inefficiencies, especially when computational or radio resources are limited.

The conservativeness of robust task offloading can be alleviated through radio resource allocation (RRA), which provides a complementary means of managing offloading latency. Since RRA adapts rapidly to varying air interface conditions, tasks' characteristics are known during RRA. By adjusting radio resources such as bandwidth based on real-time conditions, the offloading latency can be controlled. When computational resources are insufficient, additional radio resources can reduce transmission time, and vice versa, maintaining latency within acceptable bounds despite CRA uncertainty.

A key challenge in designing task offloading schemes is the mismatch in timescales among CRA, RRA, and task partitioning. CRA is long-term, task partitioning is performed for each sub-second task, and RRA operates on an even shorter timescale, adapting to dynamic wireless and mobility conditions. Effective schemes must therefore consider mixed-timescale optimization.

This paradigm aligns well with the architecture of open radio access network (O-RAN), where optimization functions are split between a non-real-time RAN intelligent controller (non-RT RIC) and a near-RT RIC \cite{polese2023understanding}. O-RAN offers a flexible, scalable platform suited for edge computing due to distributed architecture and virtualization \cite{filali2023communication}. By leveraging open interfaces, O-RAN enables disaggregated functions across distributed units (DUs) and central units (CUs), supporting seamless coordination of network resources \cite{polese2023understanding}. This distributed architecture enhances dynamic radio and computational resource allocation close to the user, a critical feature for latency-sensitive vehicular applications.

The non-RT RIC operates with latency higher than one second and handles large-timescale tasks such as policy management, system configuration, and large-scale optimization. In contrast, the near-RT RIC operates within one second to ten millisecond, making it suitable for near-instantaneous adjustments, including scheduling policy and medium access management. It processes fast-changing wireless conditions such as mobility and interference, ensuring efficient spectrum utilization and high-quality service.

In this paper, we propose a robust, resource-efficient, and latency-constrained task offloading scheme for vehicular networks in an O-RAN architecture. Our approach leverages mixed-timescale optimization to jointly perform large-timescale CRA by the non-RT RIC, moderate-timescale task partitioning by the near-RT RIC, and small-timescale RRA by DUs, ensuring latency requirements are met while minimizing resource consumption. The scheme handles uncertainty in task characteristics, making it suitable for dynamic vehicular applications where perfect knowledge of computational demands is not feasible. We provide a mathematical latency model and formulate the offloading problem as an optimization problem under a deterministic polyhedral uncertainty set. This problem is reformulated into convex sub-problems solvable using off-the-shelf solvers with polynomial complexity. The resulting solutions are reliable, near-optimal, and interpretable—key properties for safety-critical, time-sensitive applications. Moreover, optimization models do not require large training datasets and can incorporate hard constraints directly, ensuring feasibility and robustness. Learning models, in contrast, often suffer from convergence issues, optimality gaps, instability, and limited generalization. Hence, we avoid learning-based methods in favor of optimization tools offering predictable real-time performance. Results show that our robust scheme incurs no outage when demands lie in the uncertainty set, compared to about $50\%$ outage in the non-robust scheme, and our heuristic consumes about $40\%$ less bandwidth than the conventional non-slotted robust solution.

The main contributions of the paper are as follows:
\begin{itemize}
\item We develop a framework that decomposes the resource allocation problem into three timescales: small-timescale RRA adapting to channel variations, moderate-timescale task partitioning distributing computation between vehicles and edge servers, and large-timescale CRA accounting for slow VM setup and reconfiguration.

\item We propose a mixed-timescale task offloading framework compliant with O-RAN. Large-timescale CRA is performed by the non-RT RIC, moderate-timescale task partitioning by the near-RT RIC, and BS schedulers handle small-timescale RRA.

\item Since task characteristics are unknown during large-timescale CRA, we adopt worst-case robust optimization to compensate for this uncertainty. The scheme ensures latency requirements even under worst-case task data sizes, protecting network performance from variations that would otherwise degrade service quality.

\item While worst-case robust methods are often conservative, we show that integrating small-timescale RRA and moderate-timescale task partitioning effectively compensates for this conservativeness, avoiding unnecessary resource over-provisioning while satisfying latency constraints.

\item We study an important trade-off between CRA and RRA. By balancing these resource domains, the framework gains versatility in adapting to unfavorable network conditions where computational or radio resources may be lacking.
\end{itemize}}

\textcolor{black}{The paper is organized as follows. Section \ref{sec:system_model} describes the system model. Section \ref{sec:problem_statement} formulates the mixed-timescale optimization problems along with design principles. Section \ref{sec:robust_cra} presents our method for large-timescale robust CRA, and Section \ref{sec:rra} provides the solution for RRA and task partitioning. Section \ref{sec:simulation} evaluates the proposed method through simulations. Conclusions and future works are in Section \ref{sec:conclusion}.}

\begin{table*}
\caption{Comparison of Related Works}
\label{tab:comparison}
\centering
\begin{tabular}{cccccccc}
\hline
Paper & \makecell{Time Slotted\\ Optimization} & \makecell{Offloading-CSI\\Timescale Mismatch} & \makecell{Demand\\Uncertainty} & \makecell{RRA \& CRA \\ Joint Optimization} & \makecell{Latency-Constrained \\ Optimization} & \makecell{Co-Channel \\ Interference} & \makecell{Offloading \\ Strategy}\\
\hline
\cite{cong2023latency} & Non-Slotted & \ding{55} & \ding{55} & Only CRA & \ding{55} & \ding{55} & Binary\\
\cite{sun2023bargain} & Slotted & \ding{55} & \ding{55} & Only CRA & \ding{55} & \ding{51} & Binary\\
\cite{fan2023optimizing} & Non-Slotted & \ding{55} & \ding{55} & Only CRA & \ding{55} & \ding{55} & Binary\\
\cite{chang2020offloading} & Non-Slotted & \ding{55} & \ding{51} & Only RRA & \ding{55} & \ding{55} & Binary\\
\cite{tan2022decentralized} & Non-Slotted & \ding{55} & \ding{55} & Joint RRA \& CRA & \ding{55} & \ding{55} & Pure\\
\cite{li2023road} & Slotted & \ding{55} & \ding{55} & None & \ding{55} & \ding{55} & Partial\\
\cite{ku2023uncertainty} & Non-Slotted & \ding{55} & \ding{51} & Only RRA & \ding{55} & \ding{55} & Partial\\
\cite{zhao2023meson} & Slotted & \ding{55} & \ding{55} & None & \ding{55} & \ding{51} & Binary\\
\cite{abbas2024efficient} & Non-Slotted & \ding{55} & \ding{55} & Joint RRA \& CRA & \ding{55} & \ding{51} (Fixed) & Partial\\
\cite{fofana2024intelligent} & Non-Slotted & \ding{55} & \ding{55} & None & \ding{55} & \ding{55} & Partial\\
\cite{yang2025service} & Slotted & \ding{55} & \ding{55} & Joint RRA \& CRA & \ding{55} & \ding{55} & Binary\\
This work & Slotted & \ding{51} & \ding{51} & Joint RRA \& CRA & \ding{51} & \ding{51} & Partial\\
\hline
\end{tabular}
\end{table*}

{\color{black}
\section{Related Works}
VEC has been a major research focus in recent years \cite{abedi2024low, luo2025joint, zhao2025optimized, liang2025joint}. Various task offloading schemes have been developed as key enablers of VEC \cite{cong2023latency, yang2019efficient, sun2020joint, zhang2019task, sun2023bargain, fan2023optimizing, chang2020offloading, sorkhoh2019workload, tan2022decentralized, hui2020game, tajallifar2021energy, li2023road, ku2023uncertainty, zhao2023meson, abbas2024efficient, fofana2024intelligent, huang2025multimedia, yang2025service}, with their applicability influenced by several factors. In what follows, we investigate task offloading schemes.

Task offloading can be classified into three types: (a) binary offloading, where a task is fully processed locally or entirely offloaded \cite{cong2023latency, yang2019efficient, sun2020joint, zhang2019task, sun2023bargain, fan2023optimizing, chang2020offloading}; (b) pure offloading, where the entire task is always offloaded \cite{sorkhoh2019workload, tan2022decentralized, hui2020game, tajallifar2021energy}; and (c) partial offloading, where tasks are partitioned into subtasks with some processed locally and others offloaded \cite{li2023road, ku2023uncertainty, zhao2023meson, abbas2024efficient}. This flexibility allows vehicles to balance resource usage and latency.

Latency is the most prominent performance measure in task offloading strategies. Offloading is often designed for latency minimization \cite{chang2020offloading, li2023road, ku2023uncertainty, yang2025service} or joint latency–energy minimization \cite{cong2023latency, fan2023optimizing, tan2022decentralized, zhao2023meson, abbas2024efficient}. However, such objectives may yield uneven latency distribution across tasks, i.e., some tasks receive excessive resources despite already meeting latency bounds, while others may violate their requirements.

Many offloading schemes assume perfect knowledge of task characteristics \cite{cong2023latency, sun2023bargain, fan2023optimizing, tan2022decentralized, li2023road, zhao2023meson, abbas2024efficient, fofana2024intelligent, yang2025service}, which is unrealistic in many vehicular applications. In continuous applications, vehicles repeatedly send tasks and receive responses while in motion \cite{chang2020offloading, wang2016mobile}. This interactive process means that input data, computational requirements, and network conditions vary over time, preventing perfect prior knowledge.

Another uncertainty arises from partial knowledge of task dependencies within an application. For example, \cite{ku2023uncertainty} studies a partial offloading scheme for perception fusion under an uncertain task dependency graph. The task composition depends on the number of detected objects, which varies over time and is determined by the object detection task. However, this task is itself part of the application. Thus, task partitioning and offloading decisions must be made when the VEC server lacks full knowledge of the dependency graph.

Some existing works optimize only the offloading decisions, i.e., where and/or how much to offload \cite{li2023road, zhao2023meson, fofana2024intelligent}. However, this restricts the degrees of freedom needed for performance control in highly dynamic vehicular environments. Other works incorporate network resource allocation, including radio and computation resources, into task offloading to improve versatility. Specifically, \cite{chang2020offloading, ku2023uncertainty} perform RRA jointly with offloading decisions, while \cite{cong2023latency, sun2023bargain, fan2023optimizing} integrate CRA into their schemes. Higher efficiency is expected when both RRA and CRA are jointly optimized, as in \cite{tan2022decentralized, abbas2024efficient, yang2025service}. However, none of these works investigates the potential of jointly using RRA with CRA and offloading to mitigate the conservativeness of robust optimization.

Most works rely on overly simplistic assumptions about the air interface, making them prone to performance degradation in practice. Many neglect co-channel interference \cite{cong2023latency, fan2023optimizing, chang2020offloading, tan2022decentralized, li2023road, ku2023uncertainty, fofana2024intelligent, yang2025service}, despite modern systems being interference-limited. Some works consider interference; for example, \cite{sun2023bargain} accounts for inter-user interference in non-orthogonal multiple access, but does not perform RRA and ignores inter-cell interference (ICI). In \cite{abbas2024efficient}, a fixed ICI value is assumed for each orthogonal frequency division multiple access (OFDMA) sub-carrier. Hence, an efficient RRA scheme under practical assumptions is essential.

Dual-timescale optimization is common in joint task offloading and service caching \cite{lin2024deep, liu2024joint, liu2024improved, ren2022adaptive}. Since caching policies update on a larger timescale than RRA or CRA, two-timescale designs update caching on the larger timescale and RRA/CRA on the smaller one. However, these schemes do not address the intrinsic mismatch between CRA and RRA timescales.

Several works adopt a time-slotted approach to handle time-varying vehicular networks \cite{sun2023bargain, li2023road, zhao2023meson, yang2025service}. Time slot duration is generally set to the channel coherence time, so generated tasks, CRA, and RRA are assumed constant within a slot, and all tasks must be completed inside it. However, coherence times in vehicular channels are only a few milliseconds \cite{wang2018doppler}, far shorter than task completion times. Moreover, task generation periods in many applications are a few hundred milliseconds \cite{3gppV2X, 3gppenhanced}, much larger than coherence time. Thus, realistic designs must address the mismatch among task completion time, task generation period, and channel coherence time. Large-timescale CRA should therefore be combined with moderate-timescale task partitioning and small-timescale RRA for optimal task execution.

A comparison of prominent works is summarized in Table \ref{tab:comparison}. We note that existing literature lacks a mixed-timescale task offloading scheme under demand uncertainty that also incorporates practical air interface assumptions.}
\begin{figure*} [t!]
\centering
\includegraphics[width=13cm]{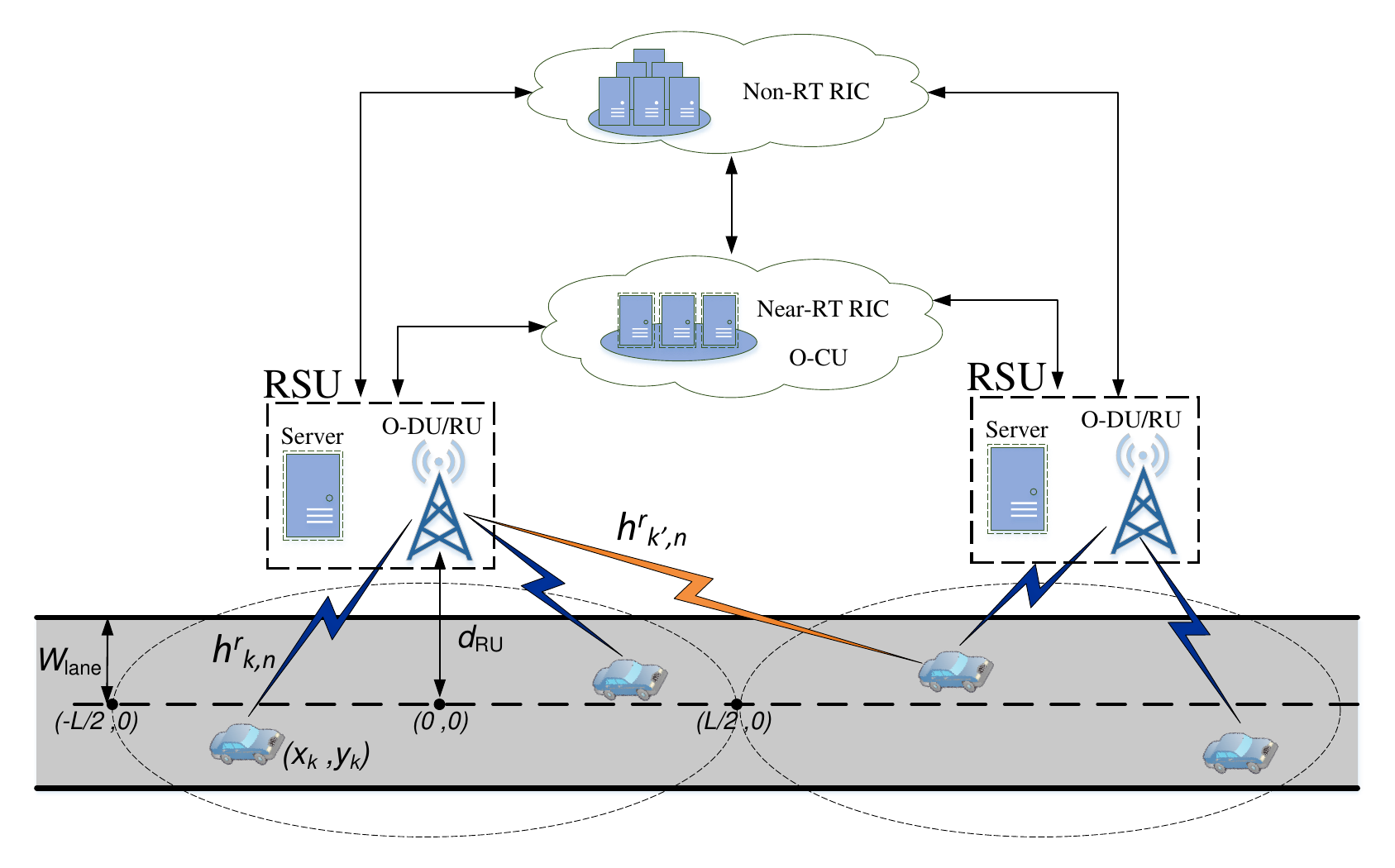}
\caption{The system model comprised of a number of RSUs controlled by a near-RT RIC and a non-RT RIC.}
\label{fig:sys_model}
\end{figure*}

\section{System Model}\label{sec:system_model}

\noindent
In this section, we describe different parts of the O-RAN architecture and their collaboration with edge servers for an efficient task offloading mechanism. \textcolor{black}{For easy reference, the notation used in the system model are summarized in Table \ref{tab:notation}.}

\begin{table}
\caption{Notations}
\label{tab:notation}
\centering
\begin{tabular}{c|c}
\hline
\bf{Parameter} & \bf{Description} \\
\hline
$\mathcal{K}$/$K$ & Set/number of vehicles\\
$R$ & Number of VEC servers/RSUs\\
$\mathcal{K}_r$& Set of vehicles under the coverage of RSU $r$\\
$\mathcal{S}_k$ & Application of vehicle $k$\\
$s_{k,j}$ & $j^{\rm th}$ task in $\mathcal{S}_k$\\
$D_{k,j}$ & data size of $s_{k,j}$\\
$\delta_k$ & Required CPU cycles per bit of $\mathcal{S}_k$\\
$D_{k}^{\rm dd}$ & Deadline of tasks in $\mathcal{S}_k$\\
$\lambda_{k,j}$ & Partitioning ratio of $s_{k,j}$\\
$T_{\rm g}$/$N_{\rm g}$ & Duration/number of task generation periods\\
$T_{\rm c}$ & Duration of large-timescale\\
$\bar{D}_{k}$ & Nominal data size\\
$\hat{D}_{k}$ & Max. deviation of data size\\
$\alpha_{k,j}$ & Uncertainty of $s_{k,j}$\\
$\Gamma$ & Uncertainty budget\\
$\mathcal{D}$ & Uncertainty set of all tasks\\
$(x_k, y_k)$ & Position of vehicle $k$\\
$N_{\rm lane}$/$N_{\rm lane}$ & Number/width of lanes\\
$L$ & Length of road segment\\
$d_{\rm RU}$/$H_{\rm RU}$ & Vertical position/height of O-RUs\\
$T_{\rm s}$/$N_{\rm s}$ & Duration/number of time slots\\
$q_{k,n}$ & Optimization variable of sub-carrier $n$ for vehicle $k$  \\
$g_{k,n}^r(t)$ & Small-scale fading between vehicle $k$ and RSU $r$\\
$d_{k,r}(mT_{\rm g})$ & Distance between vehicle $k$ and RSU $r$ on period $m$\\
$\alpha_{\rm PL}$ & path loss exponent\\
$I_{k,n}(t)$ & Interference on sub-carrier $n$ and vehicle $k$ on time slot $t$\\
$N_{\rm sc}$/$B_{\rm sc}$/$P_{\rm sc}$ & Number/bandwidth/power of each sub-carrier\\
$R_k(t)$ & Data rate of vehicle $k$ on time slot $t$\\
$Q_k(t)$ & Backlog of queue $k$ on time slot $t$\\
$f_k^{\rm vec}$ & Computational resource variable of vehicle $k$\\
$f_k^{\rm loc}$ & Computational capacity of vehicle $k$\\
$T_{k,j}^{\rm vec}$/$T_{k,j}^{\rm loc}$ & VEC server/local computation delay\\
$C_{\rm max}$ & VEC server computational capacity\\
$T_{k,j}^{\rm u}$ & Uplink transmission delay of $s_{k,j}$\\
$\xi$ & Cost weight\\
$V^{\rm tol}$ & Cutting-set tolerance\\
$\nu_k$ & Speed of vehicle $k$\\
$K_{\rm f}$ & Rician K-factor\\
$\bar{g}$ & Average of small-scale fading\\
$V_{\rm pen}$ & Penalty weight of Lyapunov optimization\\
\hline
\end{tabular}
\end{table}

\subsection{O-RAN Model}
\noindent
Fig. \ref{fig:sys_model} illustrates a set of vehicles denoted by \( \mathcal{K} = \{1, 2, \dots, K \} \) moving on a road segment and are served by $R$ RSUs. Each RSU consists of an O-DU and an O-RU, responsible for communication with the vehicles, and a VEC server that executes the vehicles' applications. Each O-DU $r$ serves a subset of vehicles $\mathcal{K}_r \subset \mathcal{K}$. O-DUs are controlled by the O-CU in O-RAN architecture, jointly collaborating on RRA and interference management among vehicles served by various O-DUs. The near-RT RIC is collocated with the CU and is responsible for moderate-timescale task partitioning. It also collects vehicle's data and passes them to a non-RT RIC for large-timescale CRA.

\subsection{Task Model}
\begin{figure} [t]
\centering
\includegraphics[width=9cm]{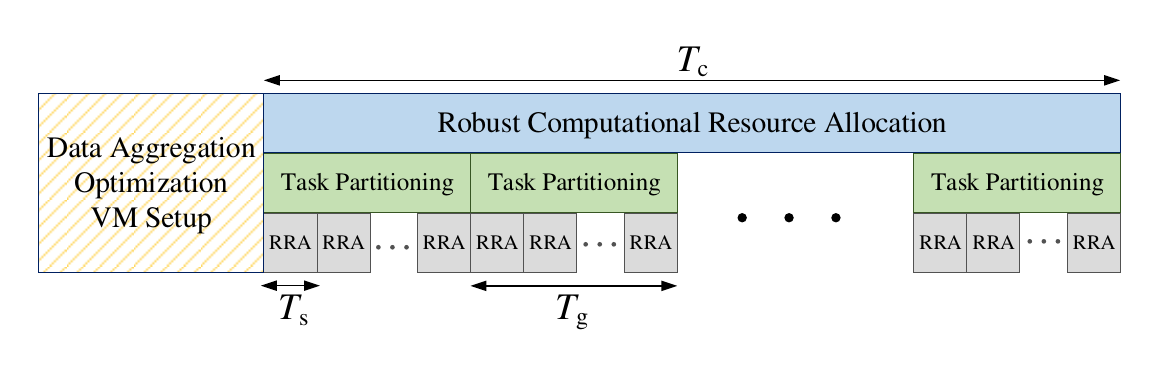}
\caption{Timeline of the system comprising $T_{\rm c}$, $T_{\rm g}$, and $T_{\rm s}$ for CRA, task generation, and RRA, respectively.}
\label{fig:timeline}
\end{figure}

\noindent
\textcolor{black}{As illustrated in Fig. \ref{fig:timeline}, the set of vehicles within the coverage area of an O-DU request vehicular applications from the associated VEC server by sending their requests via an uplink established by the O-CU and O-DU at the beginning of the large-timescale operation of the network.} Let \( \mathcal{S}_k \) represent the vehicular application of vehicle \( k \), which periodically generates a sequence of tasks \( \mathcal{S}_k = \{s_{k,1}, s_{k,2}, \dots \} \), where \( s_{k,j} \) denotes the \( j^{\text{th}} \) task generated by the application of vehicle \( k \). The task generation period is denoted by \( T_{\rm g} \), typically within the sub-second range \cite{3gppV2X, 3gppenhanced}.  Each task \( s_k \) is characterized by \( \langle D_{k,j}, \delta_k, T^{\rm dd}_k\rangle \), where \( D_{k,j} \) is the data size of input data, \( \delta_k \) is the number of required CPU cycles per each bit of data, and \(T^{\rm dd}_k\) is the deadline of tasks in vehicular application $\mathcal{S}_k$.

Each task $s_{k,j}$ is divisible into two sub-tasks, one executed locally by the vehicle $k$ and the other is offloaded and executed at the corresponding VEC server. The near-RT RIC determines the ratio $\lambda_{k,j} \in [0,1]$ of offloaded data to the entire data, and the vehicle $k$ divides the task data into a VEC server sub-task with size $\lambda_{k,j}D_{k,j}$ for offloading and a local sub-task with size $(1-\lambda_{k,j})D_{k,j}$ for local computation.

\subsection{Uncertainty Model}
\noindent
When vehicle \( k \) requests a vehicular application, the non-RT RIC allocates a portion of the computational resources of the associated VEC server to that application. Then, the VEC server configures and sets up a VM with specified resources determined by non-RT RIC. However, VM configuration is time-consuming and cannot be updated for each task within a sub-second range \cite{xia2018power, tang2019systematic}. As shown in Fig. \ref{fig:timeline}, we assume that the VEC server reconfigures the VMs every \( T_{\rm c} \) seconds, where \( T_{\rm c} \) is a design parameter, typically spanning several task generation periods \( T_{\rm g} \), i.e., \( T_{\rm c} = N_{\rm g} T_{\rm g}\) . Consequently, the exact task characteristics are unknown at the time of CRA. Nevertheless, the practical range of values for the task characteristics can be estimated based on the type of application. Specifically, we assume that the data size of each task lies within a polyhedral uncertainty set, which is defined as follows
\begin{multline}\label{eq:uncertainty_set}
\mathcal{D} = \{D_{k,j} = \bar{D}_{k} + \alpha_{k,j} \widehat{D}_{k},\quad \alpha_{k,j} \in [-1, 1], \forall k,j;\\
\quad \sum_{k=1}^K |\alpha_{k,j}| \le \Gamma, \forall j\},
\end{multline}
where $\bar{D}_{k}$ is the nominal value of data size of tasks in application $\mathcal{S}_k$, $\widehat{D}_{k}$ is the maximum deviation of data size of tasks in application $\mathcal{S}_k$ from its nominal value, and $\Gamma \in [0, K]$ determines the uncertainty budget in data size of tasks in application $\mathcal{S}_k$. In other words, $\Gamma$ determines the maximum number of tasks that can maximally deviate from their nominal values for all vehicles in the network. In particular, when $\Gamma = K$, the uncertainty set $\mathcal {D}$ is a box with the highest level of uncertainty. In contrast, there is no uncertainty when $\Gamma = 0$.

\textcolor{black}{The uncertainty in data sizes makes the required CPU cycles uncertain. In fact, since the total required CPU cycles for task $s_{k,j}$ is $\delta_k D_{k,j}$, the deviation in total required CPU cycles is $\delta_k \alpha_{k,j}\widehat{D}_{k}$ according to \eqref{eq:uncertainty_set}. The value of $\delta_k$ mostly depends on the type of vehicular application and can be known in advance. However, an uncertainty set similar to $\mathcal{D}$ \eqref{eq:uncertainty_set} can be defined on $\delta_k$ in cases where the number of CPU cycles per bit is also uncertain. Extending the solutions provided in the subsequent sections to uncertain $\delta_k$ is relatively straightforward. Therefore, we assume certain $\delta_k$ in the sequel for a simpler presentation.}

The uncertainty in input data size directly translates to uncertainty in the required CPU cycles. Specifically, since the total number of CPU cycles for task $s_{k,j}$ is given by $\delta_k D_{k,j}$, its deviation is $\delta_k \alpha_{k,j} \widehat{D}_{k}$ according to \eqref{eq:uncertainty_set}. The parameter $\delta_k$, representing the number of CPU cycles per bit, primarily depends on the vehicular application type and is typically known a priori. Nevertheless, in scenarios where $\delta_k$ is also uncertain, an uncertainty set analogous to $\mathcal{D}$ in \eqref{eq:uncertainty_set} can be defined. Extending the proposed framework to accommodate uncertainty in $\delta_k$ is conceptually straightforward and follows the same methodology as that presented for $D_{k,j}$. Therefore, for clarity of exposition, $\delta_k$ is assumed to be deterministic in the remainder of this paper.


\subsection{Vehicle Motion Model}
\noindent
According to Fig. \ref{fig:sys_model}, each RSU covers a road segment of length \( L \) with $N_{\rm lane}$ lanes each with $W_{\rm lane}$ width. Each vehicle is equipped with a global positioning system receiver to extract the corresponding location and mobility information on the road segment.  \textcolor{black}{On task generation period $m$, vehicle $k$ moves with velocity $\nu_k(mT_{\rm g})$ and its position is represented by $(x_k(mT_{\rm g}), y_k(mT_{\rm g}))$. The trajectory of vehicle $k$ is denoted by $\mathcal{T}_k = \{(x_k(mT_{\rm g}), y_k(mT_{\rm g}))), \forall m\}$, which means that we assume constant location during sub-second intervals of length $T_{\rm g}$. Trajectories are either planned by vehicles and declared to the non-RT RIC at the time of offloading request, or they can be predicted by the non-RT RIC via data-driven approaches. The O-RUs are located at the coordinates \( (x_r, d_{\rm RU}) \). The distance between vehicle \( k \) and the RSU $r$ can be known from vehicles' trajectory and is assumed to be constant within each $T_{\rm g}$ seconds. The distance is given by \( d_{k,r}(mT_{\rm g}) = \sqrt{(x_k(mT_{\rm g}) - x_r)^2 + (y_k(mT_{\rm g}) - d_{\rm RU} )^2 + H_{\rm RU}^2} \), where \( H_{\rm RU} \) is the height of the O-RU.}

\subsection{Communication Model}
\noindent
Upon generation of task \( s_{k,j} \), vehicle \( k \) transmits the associated sub-task to the RSU $r$ via uplink transmission. We employ OFDMA for the uplink. The coherence time of wireless channels in a typical vehicular environment is typically about ten milliseconds \cite{wang2018doppler}, which is considerably smaller than $T_{\rm g}$ for many vehicular applications \cite{3gppV2X, 3gppenhanced}. As a result, we divide each task generation period \( T_{\rm g} \) into $N_{\rm s}$ time slots of length \( T_{\rm s} \), smaller than the coherence time of the channel. Note that the uploading time of a task is typically larger than $T_s$. For example, for uploading a small task of size $100$ KB, a vehicle with a moderate data rate of $10$ Mbps needs $80$ ms time, which is much larger than the coherence time of wireless channels in vehicular networks. As a result, we deviate from existing works that perform only one RRA for uploading a task. Instead, we perform several RRAs until task uploading is complete.

In each time slot, the O-CU/O-DUs allocate sub-carriers to vehicle \( k \) served by RSU $r$.
We define the binary variable \( q_{k,n}(t) \), which indicates whether sub-carrier \( n \) is allocated to vehicle \( k \) in time slot \( t \).
{\color{black}\subsubsection{Channel Model}
The channel gain between each vehicle $k$ and RSU $r$ on sub-carrier $n$ in time slot $t$ is consisted of a small-scale fading component $g_{k,n}^r(t)$ and a large-scale path loss $d_{k,r}^{-\alpha_{\rm PL}}(mT_{\rm g})$, where \( \alpha_{\rm PL} \) is the path-loss exponent.
\subsubsection{Interference Model}
Each sub-carrier in OFDMA is allocated to at most one vehicle in each O-DU coverage. Hence, the uplink transmission is free of intra-cell interference. However, vehicles in neighboring cells that are assigned the same sub-carrier would introduce inter-cell interference to the uplink transmission. In particular, vehicle $k$ experiences the interference $I_{k, n}(t)$ on sub-carrier $n$ from all vehicles $k'$ in neighboring cells, where $I_{k, n}(t) = \sum_{k'\in \mathcal{K}/\mathcal{K}_r}P_{\rm sc}q_{k',n}(t)d_{k',r}^{-\alpha_{\rm PL}}(mT_{\rm g})g_{k',n}^r(t)$, and \(P_{\rm sc}\) is the per sub-carrier transmit power. Note that the interference is affected by the sub-carrier allocation in neighboring cells $q_{k',n}(t)$.

\subsubsection{Rate Expression}
According to the channel and interference models, the} achievable rate for vehicle \( k \) in time slot \( t \) is expressed as
{\color{black}\begin{equation}\label{eq:inst_rate}
R_k(t) =\sum_{n = 1}^{N_{\rm sc}} B_{\rm sc} \log_2 \bigg( 1 + \frac{P_{\rm sc}q_{k,n}(t) d_{k,r}^{-\alpha_{\rm PL}} g_{k,n}^r(t)}{N_0 B_{\rm sc} + I_{k, n}(t)} \bigg),
\end{equation}
where \( B_{\rm sc} \) is the sub-carrier bandwidth, \( N_{\rm sc} \) is the number of sub-carriers, and \( N_0 \) is the noise power spectral density.

\subsubsection{Sub-carrier Constraints}
The} total number of allocated sub-carriers in each time slot and each RSU must not exceed the total available sub-carriers, i.e.,
\begin{equation} \label{eq:carrier}
\sum_{ k\in \mathcal{K}_r} \sum_{n = 1}^{N_{\rm sc}} q_{k,n}(t) \leq N_{\rm sc}, \quad \forall r, \forall t.
\end{equation}
Besides, to avoid the intra-cell interference, we ensure that each sub-carrier is allocated to at most one vehicle in each cell. Thus, we have
\begin{equation} \label{eq:carrier_cell}
\sum_{k \in \mathcal{K}_r}  q_{k,n}(t) \leq 1, \quad \forall n,  \forall r, \forall t.
\end{equation}
\subsection{Queuing Model}\noindent
In each time slot, only a portion of tasks' data can be uploaded to O-DUs. Specifically, the data transferred from vehicle \( k \) to the serving O-DU in time slot \( t \) is \( T_{\rm s} R_k(t)\), which is generally smaller than $\lambda_{k,j}D_{k,j}$. Therefore, the remaining data of each sub-task should be uploaded in subsequent time slots. This situation resembles a queuing system with $K$ queues with backlog $Q_k(t)$ for vehicle $k$ in time slot $t$. The queue evolves as
\begin{multline}\label{eq:actual_queue}
\hspace{-9pt}
Q_k(t+1) = \max \{Q_k(t) - T_{\rm s} R_k(t) + \mathbbm{1}(t = jT_{\rm g})\lambda_{k,j}D_{k,j}, 0\}, \\
\forall t,\forall j \in \{0,1,\cdots\},
\end{multline}
where $\mathbbm{1}(.)$ is an indicator function that returns one when the argument is true and returns zero otherwise. Note that the data arrival occurs every $N_{\rm s}$ time slots upon generation of each task.

\subsection{Computation Model}
\noindent
Once the VEC server sub-task is uploaded to the O-DU, it is forwarded to the VEC server and is executed on the associated VM with computational resources denoted by \( f_k^{\rm vec} \), which is allocated to the VM in large-timescale by the non-RT RIC. Therefore, the corresponding execution delay of task $s_{k,j}$ is \( T_{k,j}^{\rm vec} = \lambda_{k,j}\frac{\delta_k D_{k,j}}{f_k^{\rm vec}} \). The corresponding CRA must satisfy
\begin{equation} \label{eq:capacity}
\sum_{k\in \mathcal{K}_r} f_k^{\rm vec} \leq C_{\rm max}, \quad \forall r,
\end{equation}
where \( C_{\rm max} \) is the total computational capacity of the VEC servers.
The local sub-task, on the other hand, is processed at the vehicle $k$ by its onboard unit with computational capacity $f_k^{\rm loc}$. Therefore, the latency of local computation is $ T_{k,j}^{\rm loc} = \left(1-\lambda_{k,j}\right)\frac{\delta_k D_{k,j}}{f_k^{\rm loc}}$.

\subsection{Latency Model}
\noindent
The total latency of task offloading is comprised of uplink transmission latency $T^{\rm u}_{k,j}$ and the VEC server sub-task execution latency $T^{\rm vec}_{k,j}$. The uplink transmission latency is dependent on the number of time slots required to finish the transmission of entire sub-tasks' data, and can be characterized as
\begin{equation} \label{eq:uplink}
T^{\rm u}_{k,j} = \sum_{t = jN_{\rm s}}^{(j+1)N_{\rm s}-1} T_{\rm s}\gamma_k(t), \forall k,
\end{equation}
where $\gamma_k(t) = \min \{\frac{Q_k(t)}{T_{\rm s}R_k(t)} , 1 \}$, i.e., if the size of the queue backlog is larger than the offered data rate in time slot $t$, one time slot length is added to the uplink transmission latency. Otherwise, the added latency would be proportional to $\frac{Q_k(t)}{T_{\rm s}R_k(t)}$.

Upon completion of the VEC server sub-task execution, the VEC server sends the processed result to the O-DU for downlink transmission. \textcolor{black}{Many applications in autonomous driving or advanced driver assistance systems hold asymmetric input-output data volume property. Specifically, these applications input large amounts of data from cameras, LiDAR, and other sensors, but produce small data size outputs such as control commands, object labels, bounding boxes, braking commands, and driver alerts \cite{baidya2020vehicular, musa2021review, tatar2024recent}.
} Additionally, the power budget of the O-RU is substantially higher than that of the vehicles. Therefore, the downlink transmission delay of the output data to the vehicles is considered negligible. Consequently, the total experienced delay of task offloading is given by \( T^{\rm u}_{k,j} + T^{\rm vec}_{k,j} \).

\textcolor{black}{We note that our mixed-timescale task offloading method can readily be modified for applications like augmented reality that may not hold the input-output data volume asymmetry. In such cases, the overall resource allocation strategy remains intact, but the latency model would incorporate the downlink delay, which depends on sub-carrier allocation in small timescale. }

Each task is considered executed when both the VEC server sub-task and the local sub-task are executed. Thus, the total task execution latency of the task $s_{k,j}$ is $\max \, \{ T^{\rm u}_{k,j} + T^{\rm vec}_{k,j}, \, T^{\rm loc}_{k,j}\}$.
\subsection{Cost Model}\noindent
It is crucial for a service provider to serve the vehicles by consuming the lowest possible resources. The network-wide resources in the proposed task offloading framework include consumed bandwidth at the air interface in the uplink transmission and the computational resources at the VEC servers. Therefore, we model the resource consumption as the combination of these two resource types with proper weights, formulated by 
\begin{equation}\label{eq:objective_resource}
\mathcal{C}(t) = \xi\sum_{k = 1}^K \sum_{n = 1}^{N_{\rm sc}}\, q_{k,n}(t) + (1-\xi)\sum_{k = 1}^K f_k^{\rm vec}.
\end{equation}

\section{\textcolor{black}{Problem Statement \& Design Principles}\label{sec:problem_statement}}
\noindent
Our objective is to determine the optimal RRA every $T_{\rm s}$ seconds, the optimal task partitioning ratio every $T_{\rm g}$ seconds, and the optimal CRA every \( T_{\rm c} \) seconds, by minimizing the resource consumption while keeping the worst-case latency of all tasks under the deadline in the presence of demand uncertainty.
The optimization problem is formulated as follows
\begin{equation}\label{eq:joint_problem}
    \begin{aligned}
\underset{\{q_{k,n}(t)\}, \{f_k^{\rm vec}\},\{\lambda_k(t)\}}{\min} \qquad\qquad\qquad\qquad\qquad\qquad\qquad &\hspace{-10em}\mathcal{C}(t) \\        
\text{subject to}\qquad\qquad\qquad\qquad\qquad\qquad\qquad\qquad&\\\qquad\begin{cases}
\max \, \{ T^{\rm u}_{k,j} + T^{\rm vec}_{k,j}, \, T^{\rm loc}_{k,j}\}\le T_k^{\rm dd}, \, \forall k,j, \,\forall \{D_k\} \in \mathcal{D}\\
 \eqref{eq:carrier}, \eqref{eq:carrier_cell}, \eqref{eq:capacity},
\end{cases}
    \end{aligned}
\end{equation}

\noindent
where $\lambda_k(t) = \lambda_{k,j}, \forall t \in [jN_{\rm s}, (j+1)N_{\rm s}-1]$, which means that the task partitioning variable is fixed in each time interval of length $N_{\rm s}$ time slots, but changes over different time intervals. The problem  \eqref{eq:joint_problem} involves a large-timescale CRA by the non-RT RIC, moderate-timescale optimization of task partitioning by the near-RT RIC, and small-timescale RRA by the O-CU/O-DU. To overcome this mixed-timescale issue, the problem \eqref{eq:joint_problem} is solved by first solving a large-timescale optimization problem and then solving a joint moderate- and small-timescale optimization problem, as discussed in the following.

\subsection{Large-Timescale Optimization}

\noindent
The non-RT RIC performs CRA at large-timescale every \( T_{\rm c} \) seconds. However, CRA depends on the moderate-timescale task partitioning and small-timescale RRA, which, in turn, are influenced by fast-varying channel conditions, typically unknown at the time of CRA.
To resolve this issue, we replace instantaneous uplink transmission rates with ergodic rates, which are independent of small-timescale channel state information (CSI). Note that while instantaneous rates may occasionally be lower than the ergodic rate, it is unlikely to lead to latency violations, as task partitioning and RRA are designed for the worst-case data size of tasks. Besides, using the worst-case rate along with worst-case data size would result in overly conservative RRA and CRA.

To transform the mixed-timescale optimization problem into a large-timescale one, we define hypothetic sub-carrier allocation $\bar{q}_{k,n}$ and task partitioning $\bar{\lambda}_k$ optimization variables, which are assumed constant in large-timescale. We refer to $\bar{q}_{k,n}$ and $\bar{\lambda}_k$ as hypothetic because they are computed for the expected uplink data rate and not actually be applied by O-RAN and vehicles. From \eqref{eq:inst_rate}, the ergodic uplink transmission rate is given by

\begin{multline}\label{eq:ergodic_rate}
\bar{R}_k = \sum_{n = 1}^{N_{\rm sc}} B_{\rm sc} \mathbb{E}_{g, d}\bigg\{\log_2 \bigg( 1 + \\
\frac{P_{\rm sc}\bar{q}_{k,n} d_{k,r}^{-\alpha_{\rm PL}}(mT_{\rm g}) g_{k,n}^r(t)}{N_0 B_{\rm sc}+ \sum_{k'\in \mathcal{K}/\mathcal{K}_r}P_{\rm sc}\bar{q}_{k',n}d_{k',r}^{-\alpha_{\rm PL}}(mT_{\rm g})g_{k',n}^r(t)} \bigg)\bigg\}.
\end{multline}

\noindent
Hence, the non-RT RIC solves the following large-timescale optimization problem

\begin{equation}\label{eq:large_problem}
    \begin{aligned}
    \underset{\{\bar{q}_{k,n}\}, \{f_k^{\rm vec}\}, \{\bar{\lambda}_k\}}{\min} \qquad\qquad\qquad\qquad\qquad\qquad\qquad &\hspace{-10em} \xi\sum_{k = 1}^K \sum_{n = 1}^{N_{\rm sc}}\, \bar{q}_{k,n} + (1-\xi)\sum_{k = 1}^K f_k^{\rm vec}\\        
\text{subject to}\qquad\qquad\qquad\qquad\qquad\qquad\qquad\qquad&\\\qquad\begin{cases}
\bar{\lambda}_k \left(\frac{D_k}{\bar{R}_k} + \delta_k \frac{D_k}{f_k^{\rm vec}}\right) \le T_k^{\rm dd}, \, \forall k, \,  \{D_{k}\} \in \mathcal{D}\\
 \left(1 - \bar{\lambda}_k\right) \delta_k \frac{D_k}{f_k^{\rm loc}} \le T_k^{\rm dd}, \, \forall k, \, \{D_{k}\} \in \mathcal{D}\\
 \eqref{eq:carrier}, \eqref{eq:carrier_cell}, \eqref{eq:capacity}.
    \end{cases}
    \end{aligned}
\end{equation}

\noindent
Once the non-RT RIC solves the large-timescale problem \eqref{eq:large_problem}, the CRA results, \( f_k^{\rm vec} \), are sent to the near-RT RIC and the VEC server, where running VMs are reconfigured or new VMs are initiated if required.
\subsection{Moderate- and Small-Timescale Optimization}\noindent
With the large-timescale computational resource values in hand, the near-RT RIC determines task partitioning every $N_{\rm s}$ time slots and sends the results to O-DU/O-CU for small-timescale RRA in each time slot. Each vehicle provides its task characteristics in its request message every $N_{\rm s}$ time slots. Consequently and in spite of large-timescale optimization, the near-RT RIC uses the certain information for optimization. Unlike prior works on task offloading, which assume constant CSI during uplink transmission, we acknowledge that the uplink transmission period typically exceeds the channel coherence time. Hence, RRA must be performed several times during uplink transmission. Similar to large-timescale problem \eqref{eq:large_problem}, we transform the latency constraint in \eqref{eq:joint_problem} into two latency constraints, and write the moderate- and small-timescale problem as 
\begin{equation}\label{eq:mixed_problem}
    \begin{aligned}
    \underset{\{q_{k,n}(t)\}, \{\lambda_k(t)\}}{\min} \quad & \sum_{k = 1}^K \sum_{n = 1}^{N_{\rm sc}}\, q_{k,n}(t) \\        
 \text{subject to} \quad & \begin{cases} 
T^{\rm u}_{k,j} + T^{\rm vec}_{k,j} \le T_k^{\rm dd}, \quad \forall k , j\\
 T^{\rm loc}_{k,j} \le T_k^{\rm dd}, \quad \forall k , j\\
\eqref{eq:carrier}, \eqref{eq:carrier_cell}.
\end{cases}
    \end{aligned}
\end{equation}

\noindent
The near-RT RIC finds the optimal value of task partitioning ratios each $N_{\rm s}$ time slots and sends the result to O-DU/O-CU, where the problem \eqref{eq:mixed_problem} is solved for $\{q_{k,n}(t)\}$ in each time slot. In addition, the optimal value of $\lambda_k(t)$ is sent to vehicles for task partitioning through a downlink control channel. \textcolor{black}{We should note that RRA in small-timescale, i.e., solving \eqref{eq:mixed_problem}, is dependent on the fast-varying small-scale fading. In contrast, CRA in large-timescale, i.e., solving \eqref{eq:large_problem}, is independent of small-scale fading because the ergodic rate in \eqref{eq:ergodic_rate} is averaged over small-scale fading distribution.}

\begin{figure} [t]
\centering
\includegraphics[width=5.5cm]{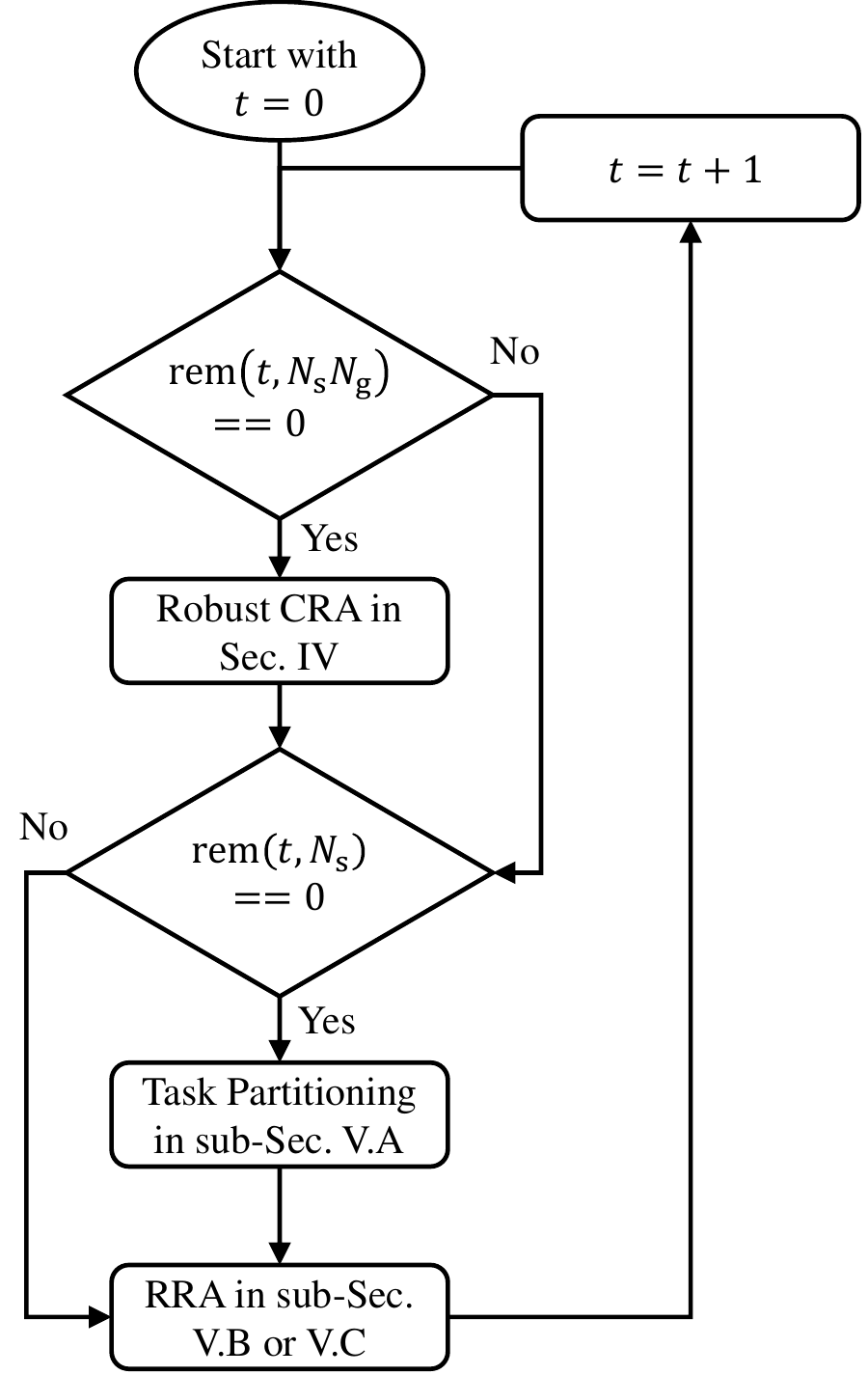}
\caption{The flowchart of the proposed scheme in system timeline.}
\label{fig:flowchart}
\end{figure}

In summary, robust CRA is performed on a large-timescale every $T_{\rm c}$ seconds, followed by moderate-timescale task partitioning every $T_{\rm g}$ seconds and small-timescale sub-carrier allocation every $T_{\rm s}$ seconds. The flowchart of the proposed scheme is illustrated in Fig. \ref{fig:flowchart}.
{\color{black}
\subsection{Signaling Flow}
We now examine the signaling flow and the corresponding overhead of the proposed framework. According to the previously described design principles, the signaling process consists of: 1) transmitting offloading requests from vehicles to the non-RT RIC in the large-timescale; 2) delivering the values of the allocated computational resources $f_k^{\rm vec}$ from the non-RT RIC to the near-RT RIC and the associated VEC servers in the large-timescale; 3) forwarding the task partitioning ratios $\lambda_k(t)$ from the near-RT RIC to the O-DU/O-CU and vehicles in the moderate-timescale; 4) transmitting task characteristics from vehicles to the near-RT RIC in the moderate-timescale; and 5) sending the indexes of the scheduled sub-carriers from O-DU/O-CU to vehicles for uplink transmission. The signaling overhead in each stage grows linearly with the number of vehicles, indicating that the proposed mixed-timescale framework is inherently scalable for large-scale vehicular networks from the perspective of signaling overhead.
}
\section{Robust CRA}\label{sec:robust_cra}
\noindent In this section we devise a scheme for the large-timescale and robust CRA by solving the problem \eqref{eq:large_problem}, which is non-convex due to coupling variables and non-convexity of the ergodic rate with respect to \( \bar{q}_{k,n}\).

The optimization problem \eqref{eq:large_problem} can be efficiently solved by the cutting-set method, which transforms the uncertain constraints to a number of corresponding constraints with certain data that are deliberately sampled from the uncertainty set \cite{mutapcic2009cutting}. Specifically, the procedure begins by solving a nominal version of \eqref{eq:large_problem} that involves only the nominal values of uncertain data sizes. Then, given the solution to the nominal problem, the worst-case data sizes are computed and the possible violation in uncertain constraints is checked. If the violation is less than a tolerable value, the robust optimization problem is solved. Otherwise, one constraint with the computed worst-case data size is added for each violated constraint. The procedure is repeated until no violation occurs. The convergence of the cutting-set method is proved in \cite{mutapcic2009cutting}. In the $i^{\rm th}$ iteration, the cutting-set method solves the following optimization problem
\begin{equation}\label{eq:cutting_set}
    \begin{aligned}
    \underset{\{\bar{q}_{k,n}\}, \{f_k^{\rm vec}\}, \{\bar{\lambda}_k\}}{\min} \qquad\qquad\qquad\qquad\qquad\qquad\qquad\quad &\hspace{-13em} \xi\sum_{k = 1}^K \sum_{n = 1}^{N_{\rm sc}}\, \bar{q}_{k,n} + (1-\xi)\sum_{k = 1}^K f_k^{\rm vec}\\
\text{subject to}\qquad\qquad\qquad\qquad\qquad\qquad\qquad\qquad&\\\qquad\begin{cases}
\bar{\lambda}_k \left(\frac{D_k}{\bar{R}_k} + \delta_k \frac{D_k}{f_k^{\rm vec}}\right) \le T_k^{\rm dd},
 \, \forall k, \, \forall\{D_{k}\} \in \widetilde{\mathcal{D}}^{(i)}_{\rm vec},\\
 \left(1 - \bar{\lambda}_k\right) \delta_k \frac{D_k}{f_k^{\rm loc}} \le T_k^{\rm dd},
 \, \forall k, \, \forall \{D_{k}\} \in \widetilde{\mathcal{D}}^{(i)}_{\rm loc},\\
 \eqref{eq:carrier}, \eqref{eq:carrier_cell}, \eqref{eq:capacity}.
    \end{cases}
    \end{aligned}
\end{equation}
\noindent
where $\widetilde{\mathcal{D}}^{(i)}_{\rm vec}$ and $\widetilde{\mathcal{D}}^{(i)}_{\rm loc}$ are the sets of worst-case data sizes for VEC server latency and local computing latency, respectively. Solving \eqref{eq:cutting_set} gives $\{\bar{q}_{k,n}^{(i)}\}, \{f_k^{{\rm vec}, (i)}\}, \{\bar{\lambda}_k^{(i)}\}$. To find the worst-case uncertain data sizes, we note that the various data sizes are dependent on each other according to the polyhedral uncertainty set in \eqref{eq:uncertainty_set}. Therefore, we find the worst-case data sizes by maximizing the summation of latency constraints of various tasks. The worst-case data sizes are obtained by
\begin{equation}\label{eq:worst-case_vec}
\{\widetilde{D}_k^{{\rm vec},(i)}\} = \underset{ \forall D_{k} \in \mathcal{D}}{\argmax} \,\sum_{k = 1}^K \, \bar{\lambda}_k^{(i)}\left( \frac{D_k}{\bar{R}_k^{(i)}} + \delta_k \frac{D_k}{f_k^{{\rm vec},(i)}}\right),
\end{equation}
where $\bar{R}_k^{(i)}$ is the ergodic rate associated with $\{\bar{q}_{k,n}^{(i)}\}$. Similarly, $\{\widetilde{D}_k^{{\rm loc}(i)}\}$ is obtained by solving 
\begin{equation}\label{eq:worst-case_loc}
\{\widetilde{D}_k^{{\rm loc},(i)}\} = \underset{ \forall D_{k} \in \mathcal{D}}{\argmax} \,\sum_{k = 1}^K \, \left(1 - \bar{\lambda}_k^{(i)}\right) \delta_k \frac{D_k}{f_k^{{\rm loc},(i)}}.
\end{equation}
The problem \eqref{eq:cutting_set} is solved when the constraint violation in both latency constraints is less than the tolerable value $V^{\rm tol}$, that is,
\begin{equation}\label{eq:vec_tolerance}
\sum_{k = 1}^K \, \bar{\lambda}_k^{(i)} \left(\frac{\widetilde{D}_k^{{\rm vec},(i)}}{\bar{R}_k^{(i)}} + \delta_k \frac{\widetilde{D}_k^{{\rm vec},(i)}}{f_k^{{\rm vec},(i)}}\right) - \sum_{k = 1}^K \, T_k^{\rm dd} \le V^{\rm tol}, \\
\end{equation}
and
\begin{equation}\label{eq:loc_tolerance}
\sum_{k = 1}^K \, \left(1 - \bar{\lambda}_k^{(i)}\right) \delta_k \frac{\widetilde{D}_k^{{\rm loc},(i)}}{f_k^{{\rm loc},(i)}} - \sum_{k = 1}^K \, T_k^{\rm dd} \le V^{\rm tol}.
\end{equation}
The cutting-set method is summarized in Algorithm \ref{alg:cutting_set}, which requires the solution to problems \eqref{eq:cutting_set}-\eqref{eq:worst-case_loc} provided in the following subsections.
\begin{algorithm}[t!]
	\label{alg:cutting_set}
	\DontPrintSemicolon
	\KwInput
	{
		$\widetilde{\mathcal{D}}^{(0)}_{\rm vec} = \{\bar{D}_k\}$, $\widetilde{\mathcal{D}}^{(0)}_{\rm loc} = \{\bar{D}_k\}$, $V^{\rm tol}$, $i = 0$
	}
	\Repeat{ Both \eqref{eq:vec_tolerance} and \eqref{eq:loc_tolerance} are true}
	{			
		$i=i+1$\\
		Solve \eqref{eq:cutting_set} and return $\{\bar{q}_{k,n}^{(i)}\}$, $\{f_k^{{\rm vec}, (i)}\}$, and $\{\bar{\lambda}_k^{(i)}\}$\\
		Solve \eqref{eq:worst-case_vec} and \eqref{eq:worst-case_loc} and return $\{\widetilde{D}_k^{{\rm vec},(i)}\}$ and $\{\widetilde{D}_k^{{\rm loc},(i)}\}$\\
		\If{\eqref{eq:vec_tolerance} is false}
		{
		$\widetilde{\mathcal{D}}^{(i)}_{\rm vec} = \widetilde{\mathcal{D}}^{(i-1)}_{\rm vec} \cup \{\widetilde{D}_k^{{\rm vec},(i)}\}$
		}
		\If{\eqref{eq:loc_tolerance} is false}
		{
		$\widetilde{\mathcal{D}}^{(i)}_{\rm loc} = \widetilde{\mathcal{D}}^{(i-1)}_{\rm loc} \cup \{\widetilde{D}_k^{{\rm loc},(i)}\}$
		}
	}
	\KwOutput{ $\{f_k^{{\rm vec}, (i)}\}$}
	
	\caption{Cutting-set method for Solving \eqref{eq:large_problem}}
\end{algorithm}
\subsection{Solving \eqref{eq:cutting_set}}
\noindent
The optimization problem \eqref{eq:cutting_set} is uncertainty-free, but still is challenging to be solved due to coupling variables and non-convexity of ergodic rate with respect to $\bar{q}_{k,n}$. To make \eqref{eq:cutting_set} tractable, we note that \( \bar{q}_{k,n}\) are \( f_k^{\rm vec} \) are coupled only in the first latency constraint. Therefore, we decouple \( \bar{q}_{k,n}\) from \( f_k^{\rm vec} \) by introducing an auxiliary variable $\varphi_k$. Specifically, we rewrite \eqref{eq:cutting_set} equivalently as
\begin{equation}\label{eq:large_decoupled}
    \begin{aligned}
    \underset{\{\bar{q}_{k,n}\}, \{f_k^{\rm vec}\}, \{\bar{\lambda}_k\}, \{\varphi_k\}}{\min}\qquad\qquad\qquad\qquad\qquad\qquad\qquad\quad &\hspace{-13em}\xi\sum_{k = 1}^K \sum_{n = 1}^{N_{\rm sc}}\, \bar{q}_{k,n} + (1-\xi)\sum_{k = 1}^K f_k^{\rm vec}\\        
\text{subject to}\qquad\qquad\qquad\qquad\qquad\qquad\qquad\qquad&\\\qquad\begin{cases}
\bar{\lambda}_k D_k \left(\varphi_k + \frac{\delta_k }{f_k^{\rm vec}}\right) \le T_k^{\rm dd}, \forall k, \, \{D_{k}\} \in \widetilde{\mathcal{D}}^{(i)}_{\rm vec},\\
\frac{1}{\bar{R}_k} \le \varphi_k, \quad \forall k,\\
\text{Other constraints in \eqref{eq:cutting_set}.}
    \end{cases}
    \end{aligned}
\end{equation}
\noindent We find the optimal task partitioning variable through the following theorem.
\begin{thm}\label{thm:lambda}
Both latency constraints of problem \eqref{eq:large_decoupled} are active at the optimal solution point $\{\bar{\lambda}_k^\star\}$, $\{\varphi_k^\star\}$, and $\{f_k^{{\rm vec}, \star}\}$.
\end{thm}
\begin{proof}
Suppose that the first latency constraint is not active at the optimal solution point for task $k$, that is,
\[
\bar{\lambda}_k^\star D_k \left(\varphi_k^\star + \frac{\delta_k }{f_k^{{\rm vec}, \star}}\right) < T_k^{\rm dd}.
\]
Hence, there is an $\widetilde{f}_k^{\rm vec} < f_k^{{\rm vec}, \star}$ such that
\[
\bar{\lambda}_k^\star D_k \left(\varphi_k^\star + \frac{\delta_k }{\widetilde{f}_k^{\rm vec}}\right) = T_k^{\rm dd}.
\]
As a result, the first latency constraint is satisfied with a lower objective value, which contradicts the optimality of $f_k^{{\rm vec}, \star}$.
Similarly, suppose that the second latency constraint is not active at the optimal solution point for task k, that is,
\[
\left(1 - \bar{\lambda}_k^\star\right) D_k \frac{\delta_k }{f_k^{\rm loc}} < T_k^{\rm dd}.
\]
Hence, there is a $\widetilde{\lambda}_k < \bar{\lambda}_k^\star$ such that the second latency constraint is active. As a result, for the first latency constraint we have
\[
\widetilde{\lambda}_k D_k \left(\varphi_k^\star + \frac{\delta_k }{f_k^{{\rm vec}, \star}}\right) < \bar{\lambda}_k^\star D_k \left(\varphi_k^\star + \frac{\delta_k }{f_k^{{\rm vec}, \star}}\right).
\]
Again, there is a $\widetilde{f}_k^{\rm vec} < f_k^{{\rm vec}, \star}$ that makes the first latency constraint active, contradicting the optimality of $\bar{\lambda}_k^\star$.
\end{proof}
According to Theorem \ref{thm:lambda}, the optimal task partitioning variable makes the latency constraint of local processing active. Therefore, we have
\begin{equation}\label{eq:lambda_large}
\bar{\lambda}_k^\star = \max\{0 , 1- \frac{T_k^{\rm dd}f_k^{\rm loc}}{ D_k \delta_k}\}.
\end{equation}
Having the optimal $\bar{\lambda}_k$ obtained, we deal with the constraint  in \eqref{eq:large_decoupled} that involves non-convex expression of the ergodic rate. To find the expected value of the achievable data rate, \textcolor{black}{we note that the channels in vehicular networks typically are with non-negligible line-of-sight (LoS) component, which is characterized by the K-factor of the Rician distribution. Particularly, empirical data in \cite{zhu2014probability} shows that the mean K-factor in urban environments is about 2 dB. K-factor values are even higher in less dense environments.} For wireless channels with dominant LoS component, K-factor is large and the probability distribution function is centered around its mean value. As a result, the ergodic rate expression  in \eqref{eq:ergodic_rate} can be approximated as follows
\begin{multline}\label{eq:rate_approx}
\bar{R}_k \approx \widetilde{R}_k = \sum_{n = 1}^{N_{\rm sc}} B_{\rm sc} \mathbb{E}_{d}\bigg\{\log_2 \bigg( 1 +\\
\frac{P_{\rm sc}\bar{q}_{k,n} d_{k,r}^{-\alpha_{\rm PL}}(mT_{\rm g}) \bar{g}}{N_0 B_{\rm sc}+ \sum_{k'\in \mathcal{K}/\mathcal{K}_r}P_{\rm sc}\bar{q}_{k',n}d_{k',r}^{-\alpha_{\rm PL}}(mT_{\rm g})\bar{g}} \bigg)\bigg\},
\end{multline}
where $\bar{g} = \mathbb{E}_g\{g_{k',n}^r(t)\} = \mathbb{E}_g\{g_{k,n}^r(t)\}$ is the average of small fading channel and is given by $\bar{g} = \sqrt{\frac{\pi}{4(K_{\rm f} + 1)}}\mathcal{L}_{1/2}(-K_{\rm f})$, and $\mathcal{L}_{1/2}(.)$ is the Laguerre function. \textcolor{black}{Although the approximation in \eqref{eq:rate_approx} may not be accurate, we note that its error is not necessarily destructive. Since we plug the mean CSI value instead of instantaneous CSI values, instantaneous data rates may happen to be below or above the ergodic rate. As a result, the large-timescale CRA may over-provision the tasks whose instantaneous rates are above the ergodic rate while under-provisions other tasks whose instantaneous rates are below the ergodic rate. In small-timescale RRA, the over-provisioned tasks demand lower number of sub-carriers, and consequently, leave more sub-carriers free to be allocated to the under-provisioned tasks, which demand higher number of sub-carriers. Therefore, the approximation error does not necessarily harm the resource consumption. We also show by computer simulations that our approximation gives relatively small errors in a wide range of K-factor values.}

The rate expression in \eqref{eq:rate_approx} is a function of the distance between vehicles and the RSU $r$, which varies every $T_{\rm g}$. \textcolor{black}{The distance depends on the mobility of vehicles, determined by their trajectory on the large-timescale. Therefore, the mobility-aware average rate over various task generation periods is}
\begin{multline}\label{eq:rate_average}
\widetilde{R}_k = \frac{1}{N_{\rm g}}\sum_{m = 0}^{N_{\rm g}-1}\sum_{n = 1}^{N_{\rm sc}} B_{\rm sc} \log_2 \bigg( 1 +\\
\frac{P_{\rm sc}\bar{q}_{k,n} d_{k,r}^{-\alpha_{\rm PL}}(mT_{\rm g}) \bar{g}}{N_0 B_{\rm sc}+ \sum_{k'\in \mathcal{K}/\mathcal{K}_r}P_{\rm sc}\bar{q}_{k',n}d_{k',r}^{-\alpha_{\rm PL}}(mT_{\rm g})\bar{g}} \bigg).
\end{multline}
\textcolor{black}{Note that we use the approximate rate in \eqref{eq:rate_average} only for large-timescale CRA, and the small-timescale RRA is performed based on the instantaneous rate with small-scale fading values.} The problem \eqref{eq:large_decoupled} is non-convex due to non-concavity of the approximate rate expression in \eqref{eq:rate_average} with respect to binary-valued sub-carrier allocation variable $\bar{q}_{k,n}$. We convexify \eqref{eq:large_decoupled} by relaxing the binary constraints to $0 \le \bar{q}_{k,n} \le 1$ and employing the convex-concave procedure (CCP), which is an iterative approach that iteratively updates a concave approximation of the rate expression \cite{lipp2016variations}. In doing so, we rewrite \eqref{eq:rate_average} as in \eqref{eq:large_rate_ccp} at the top of the next page
\begin{figure*}
	\begin{equation}\label{eq:large_rate_ccp}
	\widetilde{R}_k = \frac{1}{N_{\rm g}}\sum_{m = 0}^{N_{\rm g}-1}\sum_{n = 1}^{N_{\rm sc}} B_{\rm sc} \log_2 \bigg( N_0 B_{\rm sc} + \sum_{k'\in \{k\}\cup\mathcal{K}'_r}\bar{p}_{k',n}d_{k',r}^{-\alpha_{\rm PL}}(mT_{\rm g})\bar{g} \bigg) -\\
\underbrace{\log_2 \bigg( N_0 B_{\rm sc} + \sum_{k'\in \mathcal{K}'_r}\bar{p}_{k',n}d_{k',r}^{-\alpha_{\rm PL}}(mT_{\rm g})\bar{g}\bigg)}_{y_{k,m}(\bar{\bf p}_n)},
	\end{equation}
\hrulefill
\end{figure*}
where $\bar{p}_{k,n} = P_{\rm sc}\bar{q}_{k,n}$, $\bar{{\bf p}}_n = [\bar{p}_{1,n},\cdots, \bar{p}_{K,n}]^{\rm T}$, and $\mathcal{K}'_r = \mathcal{K}/\mathcal{K}_r$. The rate expression in \eqref{eq:large_rate_ccp} is the difference of two concave functions. Therefore, a concave approximation can be found by linearizing the second term. In $l^{\rm th}$ iteration of the CCP, the approximated rate expression as in \eqref{eq:large_rate_concave} at the top of the next page, where
\begin{figure*}
	\begin{equation}\label{eq:large_rate_concave}
\widetilde{R}_k^{(l)} = \frac{B_{\rm sc}}{N_{\rm g}}\sum_{m = 0}^{N_{\rm g}-1}\sum_{n = 1}^{N_{\rm sc}} \log_2 \bigg( N_0 B_{\rm sc} + \sum_{k'\in \{k\}\cup\mathcal{K}'_r}\bar{p}_{k',n}d_{k',r}^{-\alpha_{\rm PL}}(mT_{\rm g})\bar{g}\bigg) -\\
\bigg( y_{k,m}(\bar{{\bf p}}_n^{(l)}) + \nabla y_{k,m}(\bar{{\bf p}}_n^{(l)})^{\rm T}(\bar{{\bf p}}_n - \bar{{\bf p}}_n^{(l)}) \bigg),
	\end{equation}
\hrulefill
\end{figure*}
\begin{multline}\label{eq:large_subcarrier_grad}
\left[\nabla y_{k,m}(\bar{{\bf p}}_n^{(l)})\right]_j=\frac{\partial y_{k,m}(\bar{{\bf p}}_n^{(l)})}{\partial \bar{p}_{j,n}}= \\ \begin{cases}           
\frac{d_{j,r}^{-\alpha_{\rm PL}}(mT_{\rm g})\bar{g}}{\ln(2) \left( N_0 B_{\rm sc} + \sum_{k'\in \mathcal{K}'_r}\bar{p}_{k',n}^{(l-1)}d_{k',r}^{-\alpha_{\rm PL}}(mT_{\rm g})\bar{g} \right)},\,\forall j\in \mathcal{K}'_r,\\
0,\,\forall j \in \mathcal{K}_r,
    \end{cases}
\end{multline}
where $\ln(.)$ is the natural logarithm. The rate expression in \eqref{eq:large_rate_ccp} is a concave function, and therefore, the problem \eqref{eq:large_decoupled} with the concave approximation of the ergodic rate in \eqref{eq:large_rate_concave} is jointly convex with respect to all variables. Algorithm \ref{alg:large_ccp} summarizes the CCP for solving \eqref{eq:large_decoupled}.
\begin{algorithm}[h!]
	\label{alg:large_ccp}
	\DontPrintSemicolon
	\KwInput
	{
		$ \epsilon, T_{\rm g}$, $\widetilde{\mathcal{D}}^{(i)}_{\rm vec}$, $\widetilde{\mathcal{D}}^{(i)}_{\rm loc}$, $\bar{g}$, $d_{k,r}^{-\alpha_{\rm PL}}(mT_{\rm g})$, $\bar{\lambda}_k$, $\delta_k$, and , $\forall k,m$
	}
	Set $l = 0$\\
	\Repeat{ $\mathcal{C}^{(l-1)}(t) - \mathcal{C}^{(l)}(t) \le \epsilon$}
	{			
		$l=l+1$\\
		Compute $\left[\nabla y_k(\bar{{\bf p}}_n^{(l)})\right]$ from \eqref{eq:large_subcarrier_grad}\\
		Solve the convex approximation of \eqref{eq:large_decoupled} and return $\bar{{\bf p}}_n^{(l)}$, $\{f_k^{{\rm vec},(l)}\}, \{\bar{\lambda}_k^{(l)}\}$, and $\{\varphi_k^{(l)}\}$\\
	}
	\KwOutput{$\bar{{\bf p}}_n^{(l)}$, $\{f_k^{{\rm vec},(l)}\}, \{\bar{\lambda}_k^{(l)}\}$, and $\{\varphi_k^{(l)}\}, \, \forall n,k$.}
	
	\caption{CCP for Solving \eqref{eq:cutting_set}}
\end{algorithm}
\subsection{Solving \eqref{eq:worst-case_vec} and \eqref{eq:worst-case_loc}}
\noindent
Having the solution of \eqref{eq:cutting_set} in hand, the worst-case analysis of the cutting-set method in Algorithm \ref{alg:cutting_set} requires the solution of \eqref{eq:worst-case_vec} and \eqref{eq:worst-case_loc}. From the definition of the uncertainty set in \eqref{eq:uncertainty_set}, we rewrite \eqref{eq:worst-case_vec} as
\begin{equation}\label{eq:vec_worst-case}
    \begin{aligned}
    \underset{\alpha_k}{\max} \quad & \qquad \sum_{k = 1}^K \left(\bar{D}_k + \alpha_k \widehat{D}_k\right)\theta_k^{(i)}\\        
    \text{subject to}  & \quad\quad \begin{cases}           
-1\le \alpha_k\le 1, \forall k\\
\sum_{k = 1}^K \, |\alpha_k| \le \Gamma,\\
    \end{cases}
    \end{aligned}
\end{equation}
where $\theta_k^{(i)} = \lambda_k^{(i)}\left(\frac{1}{\bar{R}_k^{(i)}}+\frac{\delta_k}{f_k^{{\rm vec},(i)}} \right)$. The following theorem helps solving \eqref{eq:vec_worst-case}.
\begin{thm}\label{thm:worst_case}
The optimal value of $\alpha_k$ is non-negative and the uncertainty budget constraint is active at the optimal solution point.
\end{thm}
\begin{proof}
The optimal variable $\alpha_k^\star$ is non-negative because the absolute value of each feasible negative $\alpha_k$ results in larger value of the objective function due to positivity of both $\widehat{D}_k$ and $\theta_k^{(i)}$. Besides, the uncertainty budget constraint is active because if $\sum_{k = 1}^K \, \alpha_k^\star < \Gamma$, there are at least one positive $\epsilon_k$ such that $\alpha'_k = \alpha_k + \epsilon_k$ that satisfies $\sum_{k = 1}^K \, \alpha'_k = \Gamma$ and results in a higher objective value, which contradicts the optimality of $\alpha_k^\star$.
\end{proof}
According to Theorem \ref{thm:worst_case}, the problem \eqref{eq:vec_worst-case} is equivalently reformulated as
\begin{equation}\label{eq:simple_worst-case}
    \begin{aligned}
    \underset{\alpha_k}{\max} \quad & \qquad \sum_{k = 1}^K \alpha_k \widehat{D}_k\theta_k\\        
    \text{subject to} \quad & \begin{cases}           
0\le \alpha_k\le 1, \forall k\\
\sum_{k = 1}^K \, \alpha_k = \Gamma,\\
    \end{cases}
    \end{aligned}
\end{equation}
where the constant values $\bar{D}_k$ are omitted from the objective function. We find the optimal solution for the problem \eqref{eq:simple_worst-case} by the Karush–Kuhn–Tucker conditions. The Lagrangian function is
\begin{multline}
\mathcal{L}(\alpha_k, \zeta, \mu_k, \psi_k) = \sum_{k = 1}^K \alpha_k \widehat{D}_k\theta_k + \zeta\left(  \Gamma - \sum_{k = 1}^K \, \alpha_k \right) +\\
\sum_{k = 1}^K \mu_k \alpha_k + \sum_{k = 1}^K \psi_k \left(1 - \alpha_k \right),
\end{multline}
where $\zeta \ge 0$, $\mu_k \ge 0$, and $\psi_k \ge 0$ are the dual variables. The stationarity condition implies
\begin{equation}
\frac{\partial \mathcal{L}}{\partial\alpha_k} = \widehat{D}_k\theta_k - \zeta + \mu_k - \psi_k = 0.
\end{equation}
Additionally, the complementary slackness conditions are as follows
\begin{align}
\zeta\left(  \Gamma - \sum_{k = 1}^K \, \alpha_k \right) = 0,\\
\mu_k \alpha_k = 0, \quad \forall k, \\
\psi_k \left(1 - \alpha_k \right) = 0, \quad \forall k.
\end{align}
According to these conditions, we consider three possible cases for $\alpha_k$, as follows
\begin{enumerate}
\item If $\alpha_k = 1$, then $\mu_k = 0$, and the stationarity condition becomes $\widehat{D}_k\theta_k  = \zeta + \psi_k$, and $\psi_k \ge 0$ implies that $\widehat{D}_k\theta_k \ge \zeta$. Therefore, if $\alpha_k = 1$, $\widehat{D}_k\theta_k$ should be at least as large as $\zeta$.
\item If $\alpha_k = 0$, then $\psi_ k =0$, and the stationarity condition becomes $\widehat{D}_k\theta_k  = \zeta - \mu_k$, and $\mu_k \ge 0$ implies that $\widehat{D}_k\theta_k \le \zeta$. Therefore, if $\alpha_k = 0$, $\widehat{D}_k\theta_k$ should be at most $\zeta$.
\item If $0 < \alpha_k < 1$, then $\mu_k = \psi_k = 0$, and the stationarity condition reduces to $\widehat{D}_k\theta_k  = \zeta $.
\end{enumerate}
According to above, to maximize \(\sum_{k=1}^K \alpha_k \widehat{D}_k \theta_k\), the optimal solution should allocate \(\alpha_k = 1\) for as many terms as possible with the largest values of \(\widehat{D}_k \theta_k\) without violating \(\sum_{k=1}^K \alpha_k \leq \Gamma\). This approach is given in Algorithm \eqref{alg:kkt_solution}. The problem \eqref{eq:worst-case_loc} is also solved in the same manner.

\textcolor{black}{In case of uncertain $\delta_k$, problem \eqref{eq:simple_worst-case} involves maximizing a bilinear function under linear constraints, as $\theta_k$ is uncertain due to uncertainty in $\delta_k$. This problem can be simply solved by a block coordinate ascent algorithm in which a linear programming problem is solved in each iteration with respect to one uncertain parameter given the other fixed.}
\begin{algorithm}[t!]
	\label{alg:kkt_solution}
	\DontPrintSemicolon
	\KwInput
	{
		$ \Gamma, \widehat{D}_k \theta_k, \forall k$
	}
	Set $\alpha_k = 0, \, \forall k$.\\
	Sort $\{\widehat{D}_k \theta_k\}$ in descending order such that $\widehat{D}_{[1]} \theta_{[1]} \ge \widehat{D}_{[2]} \theta_{[2]} \ge \cdots \ge \widehat{D}_{[K]} \theta_{[K]}$.\\
	\For{r = 1 : K}
	{
	\If{ $\sum_{r = 1}^K \, \alpha_k \le \Gamma - 1$}
	{
	Set $\alpha_{[r] = 1}$
	}
	\ElseIf{ $\Gamma - 1  < \sum_{r = 1}^K \, \alpha_k \le \Gamma$}
	{
	Set $\alpha_{[r]} = \Gamma - \sum_{r = 1}^K \, \alpha_k$
	}
	\Else
	{
	Set $\alpha_{[r]} = 0 $.
	}
	}
	\KwOutput{ $\{\alpha_k\}$}
	
	\caption{Solving \eqref{eq:vec_worst-case}}
\end{algorithm}
\subsection{Computational Complexity (CC)}\label{sec:cra_cc}
\noindent
Here, we analyze the computational complexity (CC) of the cutting set method in Algorithm \ref{alg:cutting_set} for CRA. The CC of Algorithm \ref{alg:cutting_set} is dominated by the CC of solving \eqref{eq:cutting_set}, which is obtained from the CCP in Algorithm \ref{alg:large_ccp}. The CCP solves the convex approximation of \eqref{eq:large_decoupled}. We utilize CVX to solve \eqref{eq:large_decoupled} by exploiting iterative interior point methods (IPMs). The maximum number of iterations in IPM is $\mathcal{O}\left(\sqrt{n_{\rm c}}\log_2\left(\frac{n_{\rm c}}{\epsilon_{\rm I}}\right)\right)$, where $n_{\rm c}$ is the number of constraints and $\epsilon_{\rm I}$ is the desired accuracy \cite{boyd2004convex}. The number of constraints in \eqref{eq:large_decoupled} is $n_{\rm c} = K( 2 N_{\rm max}^{(1)} + 1) + R(N_{\rm sc} + 1)$. IPMs in each iteration solve a system of linear equations of size $n_{\rm c} + n_{\rm v}$, whose CC is of the order $(n_{\rm c} + n_{\rm v})^3$, where $n_{\rm v}$ is the number of variables. From \eqref{eq:large_decoupled}, we obtain $n_{\rm v} = K(N_{\rm sc} + 2)$. Given the maximum number of iterations in CCP, $N_{\rm max}^{(2)}$, the CC of CRA is $\mathcal{O}\left(N_{\rm max}^{(1)}N_{\rm max}^{(2)}\sqrt{n_{\rm c}}\log_2\left(\frac{n_{\rm c}}{\epsilon_{\rm I}}\right)(n_{\rm c} + n_{\rm v})^3\right)$, which is of polynomial-time complexity. \textcolor{black}{Although the CRA problem is solved every $T_{\rm c}$ seconds in a non-real-time manner, its polynomial-time complexity ensures the scalability of the large-timescale optimization in large-scale networks.}

\section{Sub-carrier Allocation And Task Partitioning}\label{sec:rra}
\noindent In this section, we propose a method for solving \eqref{eq:mixed_problem}, which involves the sub-carrier allocation in small-timescale (i.e., every time slot) and task partitioning in moderate-timescale (i.e., every $N_{\rm s}$ time slots). 
\subsection{Task Partitioning}
Every $N_{\rm s}$ time slots each vehicle generates a task and sends its characteristics to the near-RT RIC, which decides on the task partitioning portion and declares the optimal values to respective vehicles. With the same reasoning as for the large-timescale optimization, the optimal task partitioning variable is found from the fact that the latency constraint of local processing is active. Therefore, we have
\begin{equation}
\lambda_k^\star(t) = \max\{0 \, ,\, 1 - \frac{T_k^{\rm dd}(t)f_k^{\rm loc}}{D_k(t)\delta_k}\}.
\end{equation}

\subsection{Baseline Sub-carrier Allocation}\noindent
We note that \eqref{eq:mixed_problem} includes the uplink transmission delay expressed in \eqref{eq:uplink}, which requires the CSI of all time slots for uplink transmission. However, on each time slot only the CSI of the present and past time slots is available. 
A well-known scheme for solving such optimization problems is opportunistic Lyapunov optimization \cite{chai2024joint, liao2022cloud, fan2022collaborative, shi2023service}. Therefore, we use the Lyapunov optimization with certain data. Since the Lyapunov optimization operates on the problems involving time averages \cite{neely2022stochastic}, we first rewrite the problem \eqref{eq:mixed_problem} in standard time average functions as
\begin{equation}\label{eq:mixed_problem_modified}
    \begin{aligned}
    \underset{\{q_{k,n}(t)\}}{\min} \quad & \sum_{k = 1}^K \sum_{n = 1}^{N_{\rm sc}}\, q_{k,n}(t) \\        
 \text{subject to} \quad & \begin{cases} 
\sum_{t = jN_{\rm s}}^{(j+1)N_{\rm s}-1} T_{\rm s}\gamma_k(t) \le \widetilde{T}_k^{\rm dd}(t), \quad \forall k , j\\
\eqref{eq:carrier}, \eqref{eq:carrier_cell},\\
\end{cases}
    \end{aligned}
\end{equation}
where $\widetilde{T}_k^{\rm dd}(t) = T_k^{\rm dd}(t) - \lambda_k^\star(t)\delta_k\frac{D_k(t)}{f_k^{\rm vec}}$ \textcolor{black}{is the uplink delay budget of task $k$} and $D_k(t) = D_{k,j}, \forall t \in [jN_{\rm s}, (j+1)N_{\rm s}-1]$.
The Lyapunov optimization transforms optimization of time averages to per time slot optimization via opportunistic resource allocation. In doing so, a virtual queue should be defined for each latency constraint associated with uplink transmission. We denote $Z_k(t)$ as the backlog of the virtual queue associated with the uplink transmission of vehicle $k$. The backlog $Z_k(t)$ would increase when the provided uplink rate at a certain time slot is smaller than the average rate needed for maintaining the constraint and vice versa. Thus, we define the backlog $Z_k(t)$ as
\begin{multline}\label{eq:virtual_queue}
Z_k(t+1) = \max \left\{Z_k(t) - R_k(t)+ \frac{\lambda_k(t)D_k(t)}{\widetilde{T}_k^{\rm dd}(t)}, 0\right\},\\
\forall t \in \{0,1,\cdots\},
\end{multline}
Note that if the data rate in all times slots is equal to $\frac{\lambda_k(t)D_k(t)}{\widetilde{T}_k^{\rm dd}(t)}$, the latency constraint of vehicle $k$ is satisfied. As a result, the latency constraints are satisfied by keeping the backlog $Z_k(t)$ as small as possible. In other words, the latency constraints are satisfied when the virtual queues are rate stable. Lyapunov optimization ensures rate stability by transforming the problem \eqref{eq:mixed_problem_modified} to minimizing a drift-plus-penalty (DPP) function. To obtain the DPP function, the Lyapunov function is defined as a function of the queue backlogs in each time slot as follows
\begin{equation}
L(t) =  \frac{1}{2}\sum_{k = 1}^K Z^2_k(t),
\end{equation}
and the associated expected drift from time slot $t$ to $t+1$ is 
\begin{equation}\label{eq:drift}
\Delta_L(t) = \mathbb{E}\{L(t+1) - L(t)|{\mathcal Z}(t)\},
\end{equation}
where ${\mathcal Z}(t) = \{ Z_1(t), \cdots, Z_K(t) \}$. The actual queue backlogs $Q_k(t)$ are not incorporated in the Lyapunov function because there is no input data after task generation time slot and consequently, the actual queue backlogs are decreasing in each task generation period. Besides, the deadlines are supposed to be less than the task generation period $T_{\rm g}$. Therefore, each task is totally executed before generation of its subsequent task for a feasible problem. According to the problem \eqref{eq:mixed_problem_modified}, the DPP function is 
\begin{equation}\label{eq:drift}
DPP(t) = \Delta_L(t) + V_{\rm pen} \mathbb{E}\{\sum_{k = 1}^K\,  \sum_{n = 1}^{N_{\rm sc}}\, q_{k,n}(t)\},
\end{equation}
where $V_{\rm pen}$ is a design parameter that indicates the relative importance of the drift with respect to the penalty. It can be shown that an upper bound on the DPP function can be found as \cite{neely2022stochastic}
\begin{multline}\label{eq:bound}
DPP(t) \le B + \mathbb{E}\bigg\{\sum_{k = 1}^K Z_k(t) \bigg ( - T_{\rm s}R_k(t)+\\
 T_{\rm s}\frac{\lambda_k^\star(t)D_k(t)}{\widetilde{T}_k^{\rm dd}(t)}\bigg )\, \rule[-5pt]{0.7pt}{20pt} \,\mathcal{Z}(t) \bigg\} + V_{\rm pen} \mathbb{E}\bigg\{\sum_{k = 1}^K\sum_{n = 1}^{N_{\rm sc}}\, q_{k,n}(t) \, \rule[-5pt]{0.7pt}{20pt} \, \mathcal{Z}(t)\bigg\},
\end{multline}
where $B$ is a constant. It is shown that minimizing the upper bound in \eqref{eq:bound} ensures that the objective function is minimized with maximum $\mathcal{O}(1/V_{\rm pen})$ difference from the minimum achievable penalty, and the average backlog, which is proportional to the latency, is at most $\mathcal{O}(V_{\rm pen})$ larger than the minimum achievable queue backlog \cite{neely2022stochastic}. As a result, the problem \eqref{eq:mixed_problem_modified} is solved in each time slot by solving
\begin{equation}\label{eq:small_subcarrier}
    \begin{aligned}
    \underset{\{q_{k,n}(t)\}}{\max} & \quad  \sum_{k = 1}^K   R_k(t) Z_k(t)  - V_{\rm pen} \sum_{k = 1}^K\sum_{n = 1}^{N_{\rm sc}}\, q_{k,n}(t)\\
    \text{subject to} & \qquad \qquad \eqref{eq:carrier}, \eqref{eq:carrier_cell}
    \end{aligned}
\end{equation}
It is observable that when a virtual queue backlog increases, the associated weight of its rate in the objective function increases, and consequently, more bandwidth is allocated to that queue to compensate for increased backlog. As a result, solving \eqref{eq:small_subcarrier} efficiently allocates sub-carriers among various tasks to maintain the virtual queue backlogs as small as possible.

The problem \eqref{eq:small_subcarrier} is non-convex due to non-concave $R_k(t)$ and binary-valued variable $q_{k,n}(t)$. Similar to the large-timescale optimization, we achieve a convex approximation of the problem \eqref{eq:small_subcarrier} by relaxing the binary constraints to $0\le q_{k,n}(t)\le 1$, and adopting the CCP to find a concave approximation of $R_k(t)$. The details are not repeated here.

{\color{black}
Solutions to problem \eqref{eq:small_subcarrier} are relaxed continuous variables while in practice only binary variables are applicable. Thus, we need post-processing to transform the relaxed variables into binary ones. We adopt a greedy approach in which we assign sub-carrier $n$ to user $k$ when it has the highest corresponding variable among all users in the same cell and it increases the objective function. In other words, we set $q_{k,n}^{\rm bin}(t) = 1$, when $k = \argmax_{k'\in \mathcal{K}_r} q_{k',n}(t)$ in its serving RSU $r$ and we have $\sum_{k = 1}^K   R_k^{\rm bin}(t) Z_k(t)  - V_{\rm pen} \sum_{k = 1}^K\sum_{n = 1}^{N_{\rm sc}}\, q^{\rm bin}_{k,n}(t)$ strictly increasing with the new allocation, where $R^{\rm bin}_k(t)$ is the data rate resulting from binary sub-carrier allocations. Note that the value of $R_k^{\rm bin}(t)$ changes after each sub-carrier allocation due to inter-cell interference. Therefore, we should update $R_k^{\rm bin}(t)$ after each allocation to take the impact of new allocations into account. After all users examined orderly, we sweep all vehicle-sub-carrier pairs to allocate any sub-carriers left that may increase the objective function. We continue sweeping until there is no sub-carrier that increases the objective function.
}
\subsection{Heuristic Sub-carrier Allocation}\noindent
Solving \eqref{eq:small_subcarrier} in each time slot depends on the penalty value $V_{\rm pen}$. Large values of $V_{\rm pen}$ put more emphasize on lowering the bandwidth consumption, resulting in lower data rates and possible deadline violations. In contrast, small values of $V_{\rm pen}$ result in higher data rates and may avoid deadline violations. However, higher data rates may be overly conservative, resulting in excessive bandwidth consumption. This situation arises from the fact that the sub-carrier allocation in Lyapunov optimization is not explicitly dependent on the deadline values. Instead, the deadline values only impact the virtual queue backlogs as the data rate weights in the objective function of \eqref{eq:small_subcarrier}.
To resolve this issue, we propose to transform the uplink transmission constraint in \eqref{eq:mixed_problem_modified} to a per time slot data rate constraint inspired by the virtual queue evolution equation in \eqref{eq:virtual_queue}. \textcolor{black}{Specifically, we note that if the data rate of task $k$ is larger than the target rate $R^{\rm trg}_k(t) = \frac{\lambda_k(t)D_k(t)}{\widetilde{T}_k^{\rm dd}(t)}$ in each time slot $t$, the task deadline will be met. We note that the binary variables derived from the relaxed variables may either violate task deadlines or allocate extra sub-carriers. Therefore, we need to ensure that the binary variables are neither deficient nor excessive. To do so, after each time slot, we calculate any excessive or deficient data rate resulting from binarization. Particularly, $R^{\rm bin}_k(t) - R^{\rm trg}_k(t)$ determines the deficient or excess data rate due to binarization in time slot $t$. When $R^{\rm bin}_k(t) - R^{\rm trg}_k(t) \le 0$, subsequent slots should compensate for rate deficiency, i.e., we need to increase the target data rate in the subsequent time slot. Otherwise, we reduce the target data rate to spend lower resources on tasks with excessive sub-carriers. Therefore, we set the rate compensation to $\rho_k(t) = R^{\rm bin}_k(t) - R^{\rm trg}_k(t) + \rho_k(t - 1) $, initialized with $\rho_k(mT_{\rm g}) = 0$. Instead of DPP minimization, we solve the following optimization problem
\begin{equation}\label{eq:mixed_problem_heuristic}
    \begin{aligned}
    \underset{\{q_{k,n}(t)\}}{\min} \quad & \sum_{k = 1}^K \sum_{n = 1}^{N_{\rm sc}}\, q_{k,n}(t) \\        
 \text{subject to} \quad & \begin{cases} 
R_k(t) \ge R^{\rm trg}_k(t) - \rho_k(t - 1), \quad \forall k\\
\eqref{eq:carrier}, \eqref{eq:carrier_cell},\\
\end{cases}
    \end{aligned}
\end{equation}}
The problem above is non-convex, but can become convex using binary relaxation and CCP. Note that the problem \eqref{eq:mixed_problem_heuristic} does not offer a trade off between bandwidth consumption and deadline violation. Hence, as we will show by simulation results, our heuristic approach significantly outperforms the Lyapunov optimization.

{\color{black}
Similar to \eqref{eq:small_subcarrier}, solutions to problem \eqref{eq:mixed_problem_heuristic} are relaxed continuous variables and need post-processing to be transformed into binary ones. We adopt a similar greedy approach in which we assign sub-carrier $n$ to user $k$ when it has the highest corresponding variable among all users in the same cell. In contrast to binarization of \eqref{eq:small_subcarrier}, we stop allocating any sub-carriers to user $k$ when the rate constraint $R_k^{\rm bin}(t) \ge R^{\rm trg}_k(t) - \rho_k(t - 1)$ is satisfied to keep the objective of \eqref{eq:mixed_problem_heuristic} as small as possible.
}

\textcolor{black}{For applications like augmented reality that may produce large volumes of output data, we need to incorporate another small-timescale sub-carrier allocation for downlink transmission after task execution on VMs is complete. In doing so and having uplink and execution delays in hand, O-CU calculates the downlink delay budget of each task, and accordingly distributes the available sub-carriers among various tasks so that their total latency is kept under their deadlines. This can be accomplished by solving another RRA problem similar to \eqref{eq:mixed_problem_heuristic}, but with downlink delay budgets.}
\subsection{CC of RRA}\label{sec:cc_rra}\noindent
Here, we obtain the CC of RRA for both the Lyapunov optimization and our heuristic scheme. The Lyapunov optimization solves the problem \eqref{eq:small_subcarrier}, which includes \( KN_{\rm sc} \) variables and \( R(N_{\rm sc} + 1) \) constraints. Similar to the discussion in Sub-section \ref{sec:cra_cc}, we derive the CC of the Lyapunov optimization as \( \mathcal{O}\left(\sqrt{RN_{\rm sc}}\log_2 \left(\frac{RN_{\rm sc}}{\epsilon}\right)N_{\rm sc}^3 (K + R)^3\right) \), which is of polynomial-time complexity. Note that sub-carrier allocation is performed in each time slot. Thus, the overall CC is multiplied by \( \frac{T_{\rm c}}{T_{\rm s}} \), i.e., the total number of time slots in the large timescale.

Our heuristic scheme, on the other hand, solves the problem \eqref{eq:mixed_problem_heuristic} with the same number of variables as in \eqref{eq:small_subcarrier} and \( R(N_{\rm sc} + 1) + K \) constraints. We note that since \( RN_{\rm sc} \) generally dominates \( K \), the CC of our heuristic scheme is the same as that of the Lyapunov optimization despite the additional constraints in the former. \textcolor{black}{A tractable RRA is necessary for scalability of task offloading, as it is performed in a real-time manner. Accordingly, the scalability of our method in large-scale networks is guaranteed due to the polynomial-time complexity of both the Lyapunov-based and heuristic RRA methods.}

\begin{table}
\caption{Simulation Setup}
\label{tab:setup}
\centering
\begin{tabular}{cc|cc}
\hline
Parameter & Value & Parameter & Value\\
\hline
$K$ & $100$ & $N_{\rm sc}$ & $40$\\
$R$ & $10$ & $N_0$ & $-163$ dBm/Hz\\
$\bar{D}_k$ & $U(0.8,1.2)$ Mbits, $\forall k$ & \textcolor{black}{$K_{\rm f}$} & \textcolor{black}{$1$}\\
$\widehat{D}_k$ & $0.1$ Mbits, $\forall k$ & $\alpha_{\rm PL}$ & $2.5$\\
$\Gamma$ & $K, K/2$ & $L$ & $5$ Km\\
$\xi$ & $0.9$ & $N_{\rm lane}$ & $4$\\
$T_{\rm s}$ & $10$ ms & $W_{\rm lane}$ & $10$ m\\
$T_{\rm g}$ & $250$ ms & $H_{\rm RU}$ & $20$ m\\
$T_{\rm c}$ & $2$ sec & $\nu_k$ & $U(15, 25)$ m/s\\
$\sigma_{\rm x}$ & $1$ m/s & $\sigma_{\rm y}$ & $0.1$ m/s\\
$f_k^{\rm loc}$ & $2$ GHz, $\forall k$ & $P_{\rm sc}$ & $23$ dBm\\
$C_{\rm max}$ & $40$ GHz & $B_{\rm sc}$ & $1$ MHz\\
\hline
\end{tabular}
\end{table}
\section{Simulation Results}\label{sec:simulation}
\noindent In this section, we evaluate robust and non-robust methods in terms of robustness against demand uncertainty and resource consumption.

\textcolor{black}{\subsection{Setup}
We adopt the widely implemented mobility model of constant velocity with Gaussian perturbations \cite{stellet2015estimating}.} \textcolor{black}{ We assume that vehicles move at a constant velocity in each period, but velocities may change across periods. The amount of change is derived from a white Gaussian noise in both horizontal and vertical directions, respectively with standard deviations $\sigma_{\rm x}$ and $\sigma_{\rm y}$.} The parameter setting is shown in Table \ref{tab:setup}. Each data point is the average of Monte Carlo realizations.

\textcolor{black}{\subsection{Baseline Schemes}
We compare our mixed-timescale robust method with both non-slotted and non-robust methods. Specifically, we compare our robust solution based on the cutting-set method to non-robust solutions. In addition, we compare the performance of our mixed-timescale solution with non-slotted ones that allocate all resource types at the same timescale. We also compare the Lyapunov-based RRA with our heuristic method. We acknowledge that some conventional policy gradient methods in reinforcement learning (RL) are applicable to our system model. However, applying these methods necessitates devising a Markov decision process (MDP) for our system model and customizing these methods to that MDP, which entails a new set of contributions and re-organization of the paper. This can be an interesting task in our future work.}

\subsection{\textcolor{black}{Results}}
\begin{figure} [t]
\centering
\includegraphics[width=8cm]{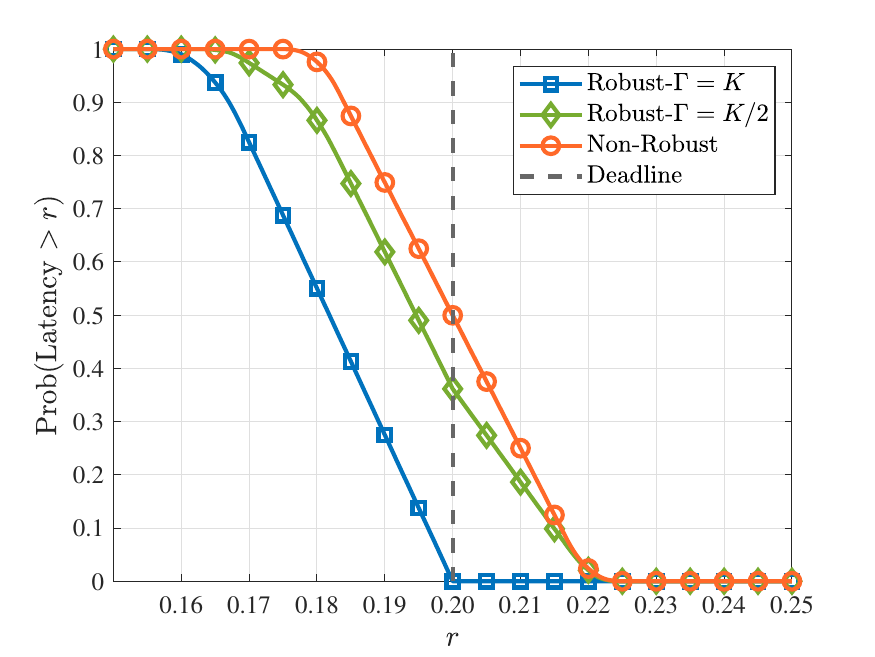}
\caption{CCDF of robust and non-robust solutions to the large-timescale optimization.}
\label{fig:ccdf_large}
\end{figure}

We first evaluate the robustness of the cutting-set method for solving \eqref{eq:large_problem}, focusing on large-timescale optimization against demand uncertainty. Fig. \ref{fig:ccdf_large} shows the complementary cumulative distribution function (CCDF) of robust and non-robust solutions. Two robust solutions are considered: one robust against a box uncertainty set with maximum uncertainty (\(\Gamma = K\)), and the other against a polyhedral set (\(\Gamma = K/2\)). In non-robust solution, we set \(\Gamma = 0\). Tasks' deadlines are set to \(0.2\) seconds, and data sizes are generated within the box uncertainty set. We observe zero outage probability (tasks' latency exceeding the deadline) for the box-robust method. The polyhedral-robust method shows some outage, as some data size realizations lie within the box set but outside the polyhedral set. The non-robust method has the highest outage probability (\(\approx 0.5\)). We also note that the robust method is conservative, achieving unnecessarily low latency for many realizations far from the worst case.

\begin{figure} [t]
\centering
\includegraphics[width=8cm]{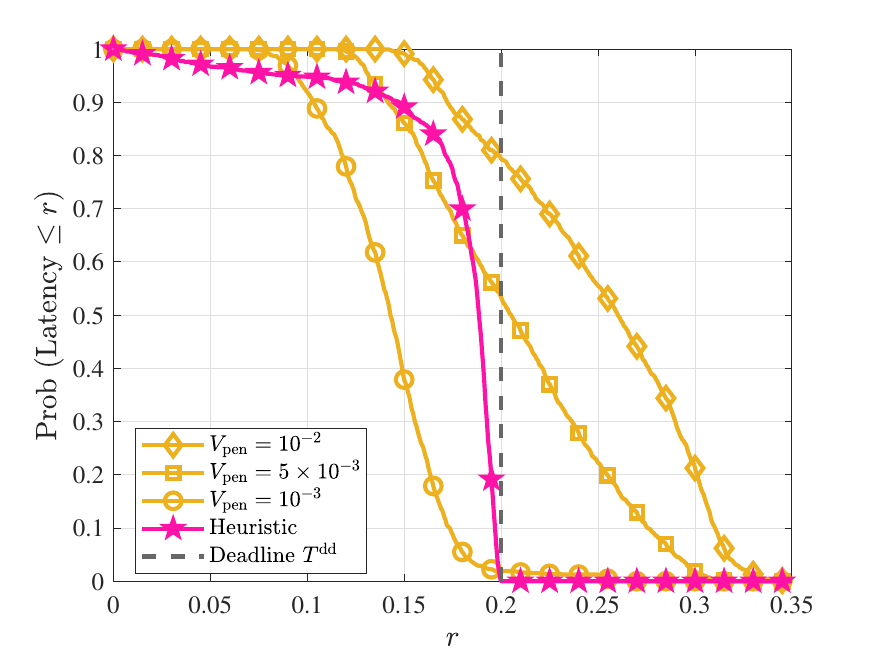}
\caption{CCDF of achieved latencies by mixed-timescale optimization methods.}
\label{fig:ccdf_small}
\end{figure}

\begin{figure} [t]
\centering
\includegraphics[width=8cm]{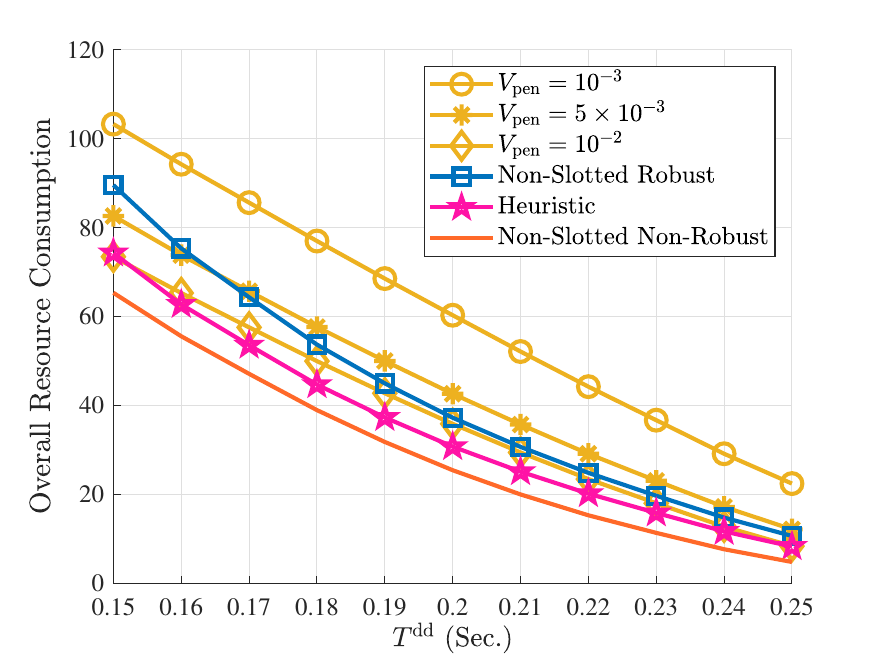}
\caption{Resource consumption vs. Tasks' deadline for various task offloading schemes.}
\label{fig:resource_small}
\end{figure}

To investigate the performance of small-timescale optimization, we show the CCDF of the achieved latency for small-timescale sub-carrier allocation in Fig. \ref{fig:ccdf_small}, which shows the achieved CCDF of the Lyapunov optimization for three different values of $V_{\rm pen}$ as well as that of our heuristic method. We observe that the Lyapunov optimization results in lower outage probabilities for smaller penalty values. It is also observable that the Lyapunov optimization is more conservative for smaller penalty values.

As discussed on Fig. \ref{fig:ccdf_large}, the large-timescale optimization results in a conservative resource allocation. In contrast, the proposed heuristic method succeeds to lower the conservatism, i.e., maintains the latency of each task closer to the deadline. By recalling that the optimal resource allocation makes latency constraints active, we conclude that our heuristic method performs similar to optimization with perfect data. In other words, the small-timescale optimization effectively mitigates the uncertainty in large timescale, which confirms our main contribution in this paper.

\textcolor{black}{We also compare the resource consumption of various task offloading schemes in Fig. \ref{fig:resource_small}. We note that our heuristic scheme outperforms the non-slotted robust scheme, which is obtained by utilizing the cutting-set method and assuming that RRA and CRA are performed simultaneously, a common assumption in the literature. Our heuristic scheme consumes less resources than the Lyapunov optimization, even for a high penalty of $V_{\rm pen} = 10^{-2}$.}

\begin{figure} [t]
\centering
\includegraphics[width=8cm]{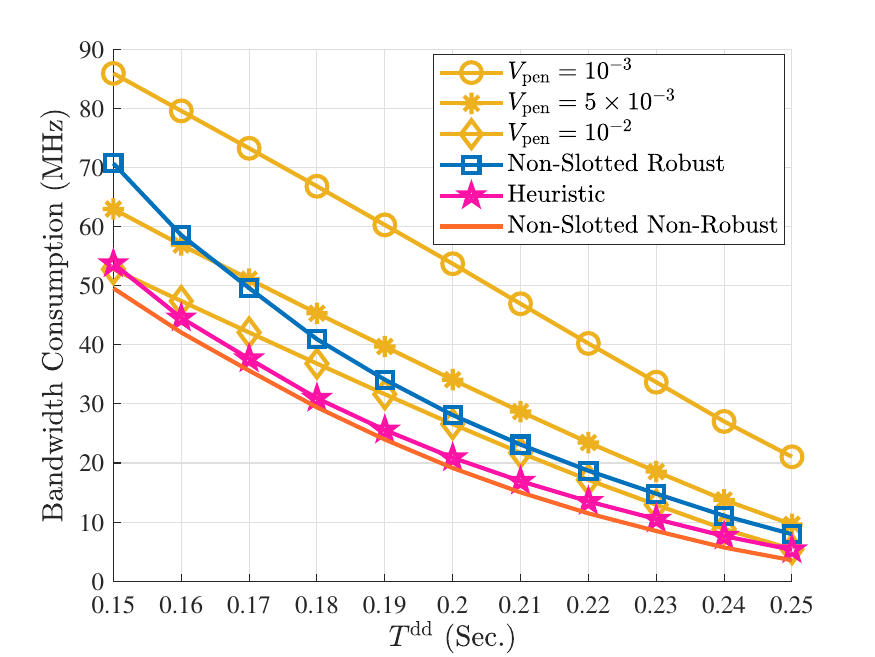}
\caption{Bandwidth consumption vs. Tasks' deadline for various task offloading schemes.}
\label{fig:bandwidth_small}
\end{figure}

\textcolor{black}{Since our heuristic scheme performs RRA differently, we show the bandwidth consumption of various robust schemes to highlight the efficiency of the proposed scheme. As shown in Fig. \ref{fig:bandwidth_small}, our heuristic scheme consumes the lowest bandwidth among robust methods. Specifically, it consumes roughly $40 \%$ less bandwidth than the Lyapunov optimization that maintains task deadlines when $V_{\rm pen} = 10^{-3}$.}

\begin{figure} [t]
\centering
\includegraphics[width=8cm]{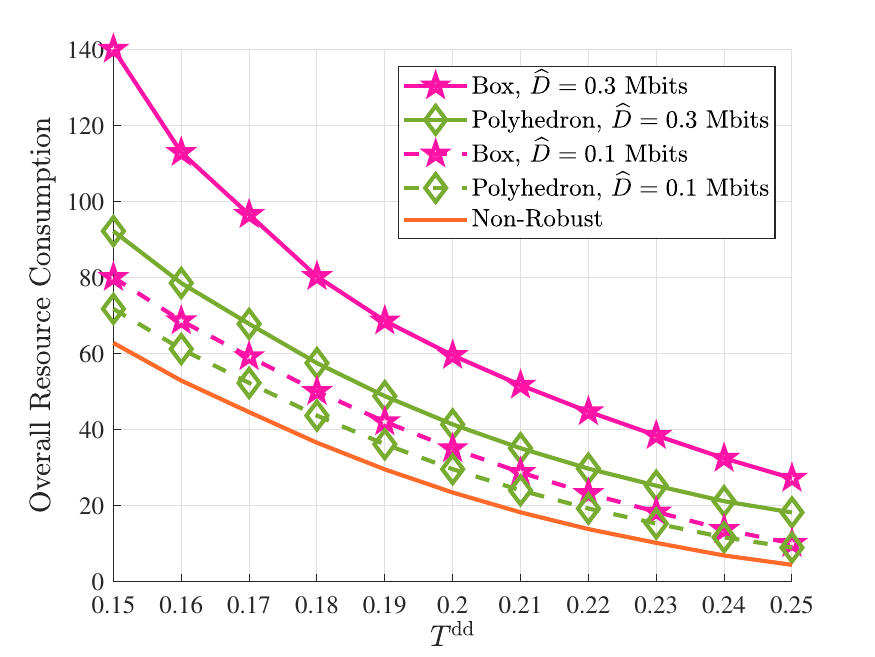}
\caption{Resource consumption of the robust and non-robust methods for mixed-timescale optimization.}
\label{fig:resource_usage}
\end{figure}

We now compare the overall resource consumption of the non-robust and robust methods. Fig. \ref{fig:resource_usage} shows the objective function of the mixed-timescale optimization problem \eqref{eq:joint_problem}. It is observable that all methods consume more resources for tighter deadlines, i.e., smaller $T^{\rm dd}$, because the feasible set of \eqref{eq:joint_problem} is smaller for smaller $T^{\rm dd}$. In addition, the resource consumption is higher for higher levels of uncertainty, i.e., either for larger deviation $\widehat{D}$ of data sizes to their nominal values or for larger $\Gamma$. Besides, the non-robust method consumes lower resources than the robust ones. As a result, the robustness comes with higher resource consumption.

\begin{figure} [t]
\centering
\includegraphics[width=8cm]{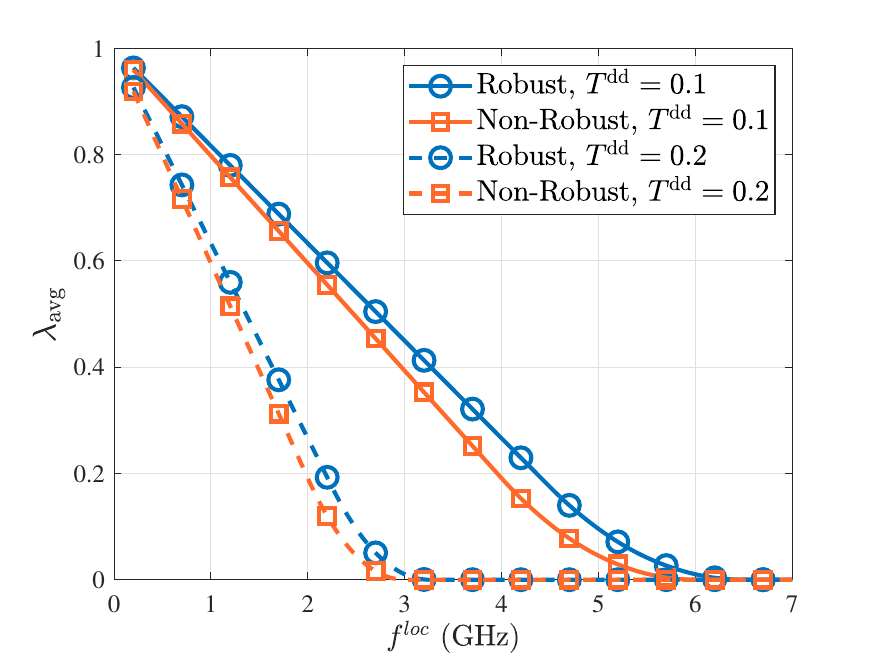}
\caption{The average of task partitioning ratio of the non-robust and robust methods.}
\label{fig:lambda_avg}
\end{figure}

We study the efficiency of the task partitioning by investigating the average of task partitioning ratio $\lambda_{\rm avg}$ for the non-robust and robust methods in Fig. \ref{fig:lambda_avg}. As shown in the figure, both methods tend to offload smaller portion of tasks for larger values of local processing capacity $f^{\rm loc}$. In other words, since offloading necessitates consumption of network-wide resources, both methods choose to execute the tasks locally, unless the deadline is violated. Besides, since the robust methods allocate resources for the worst-case tasks, the task partitioning ratio of robust methods is larger than that of non-robust ones. We also observe that the task partitioning ratio is larger for smaller deadlines because the amount of locally processed data decreases in smaller time windows, and vehicles must offload higher portions of their data to maintain the latency constraints.

\begin{figure} [t]
\centering
\includegraphics[width=8cm]{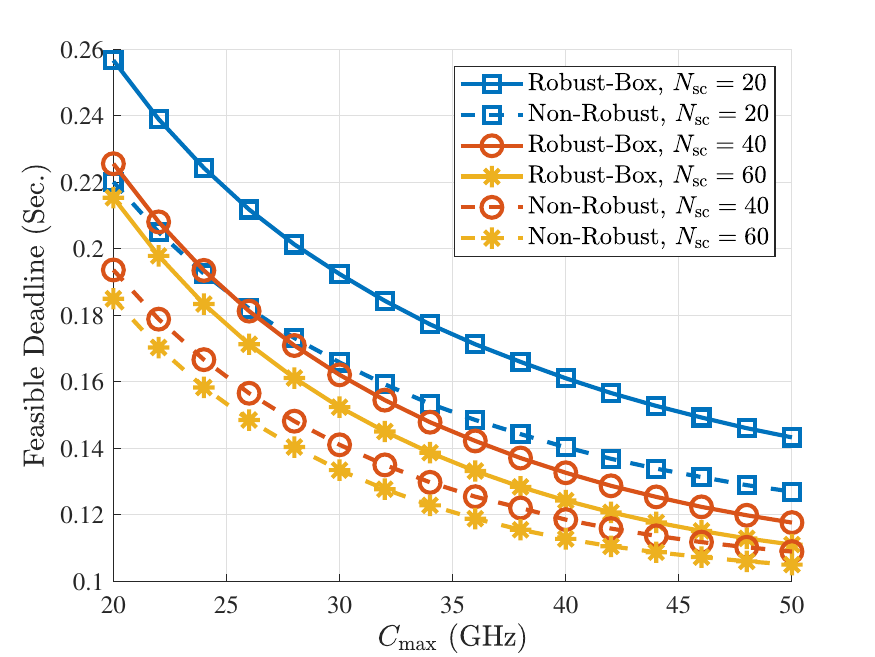}
\caption{The average of feasible deadlines with respect to computational resource capacity of RSUs and for various number of sub-carriers.}
\label{fig:deadline_cpu}
\end{figure}

We study the relation between network resource capacity and feasible deadlines in Fig. \ref{fig:deadline_cpu}. Feasible deadline refers to the lowest possible deadline that the network can serve for a given capacity in radio and computational resource domains. We obtain feasible deadlines by adding positive $\nu_k$ to $T^{\rm dd}_k$ and penalizing the objective function of \eqref{eq:joint_problem} with $\sum_{k = 1}^K \, \nu_k$. This problem is always feasible and the lowest feasible deadline of task $k$ is $T^{\rm dd}_k + \nu_k $. Fig. \ref{fig:deadline_cpu} shows that the feasible deadlines decrease for larger values of $C_{\rm max}$. Similarly, tasks with tighter deadlines can be served with higher bandwidth in air interface. Fig. \ref{fig:deadline_cpu} also reveals a trade off between consumption of radio and computational resources. Specifically, the lack of computational resources can be compensated for by adding more sub-carriers to the air interface and vice versa. This also allows a higher level of versatility for the non-RT RIC to provide more service stability in unfavorable network conditions.

\begin{figure} [t]
\centering
\includegraphics[width=8cm]{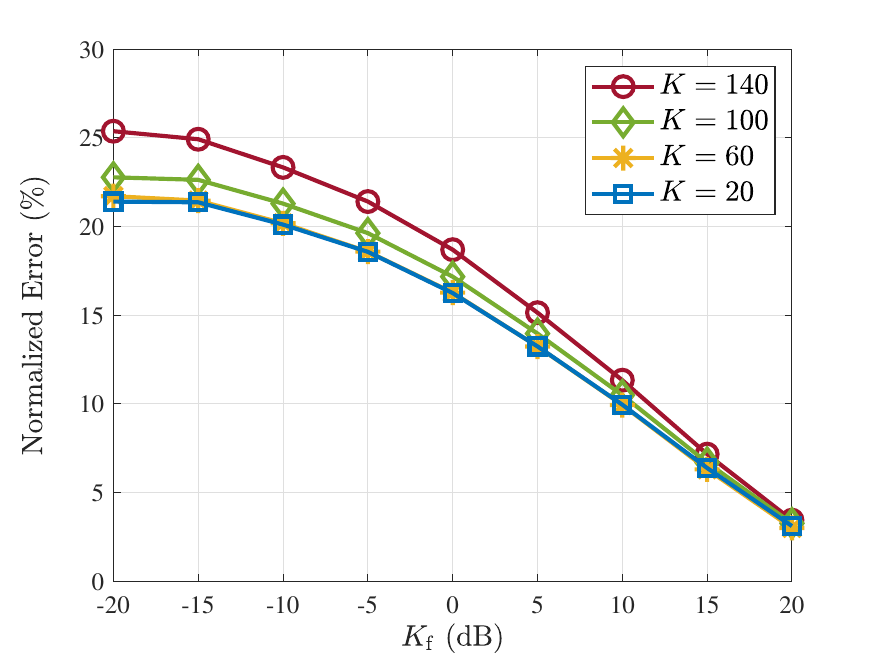}
\caption{The normalized error of ergodic rate approximation.}
\label{fig:ergodic_error}
\end{figure}

In the large-timescale optimization, we approximated the ergodic data rate. To evaluate the accuracy of the approximation, we generated many instantaneous data rates from instantaneous CSI and computed the normalized error with the ergodic rate in the right-hand side of \eqref{eq:rate_approx}. Fig. \ref{fig:ergodic_error} shows the normalized approximation error for various values of $K_{\rm f}$. As expected, the error reduces for higher values of $K_{\rm f}$, i.e., the approximation is more accurate when the Rician channel is more centered around its mean value. Besides, we note that the normalized approximation error is less than 25$\%$ for extremely small $K_{\rm f}$ values, \textcolor{black}{which shows that the approximation does not significantly deviate from the true value even in Rayleigh fading channels.}

\begin{figure} [t]
\centering
\includegraphics[width=8cm]{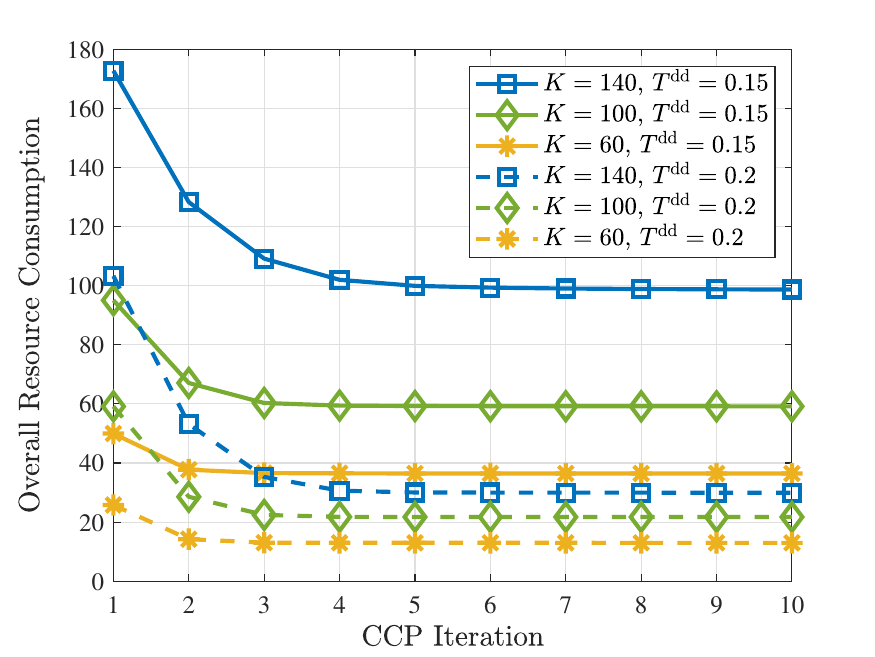}
\caption{Convergence behavior of the CCP method.}
\label{fig:ccp_iteration}
\end{figure}

\textcolor{black}{We now analyze the convergence behavior of the CCP method applied to the large-timescale CRA. Fig. \ref{fig:ccp_iteration} shows the overall resource consumption, i.e., the objective function of problem \eqref{eq:cutting_set}, across iterations of the CCP in Algorithm \ref{alg:large_ccp}. Fig. \ref{fig:ccp_iteration} shows fast convergence of CCP, as it reaches a steady value after just four or five iterations in various parameter setting of vehicles and deadlines. Since CCP is executed in each iteration of the cutting-set method, the convergence may become even faster in subsequent iterations by initiating the CCP with the obtained solutions in previous iterations. 
}

\begin{figure} [t]
\centering
\includegraphics[width=8cm]{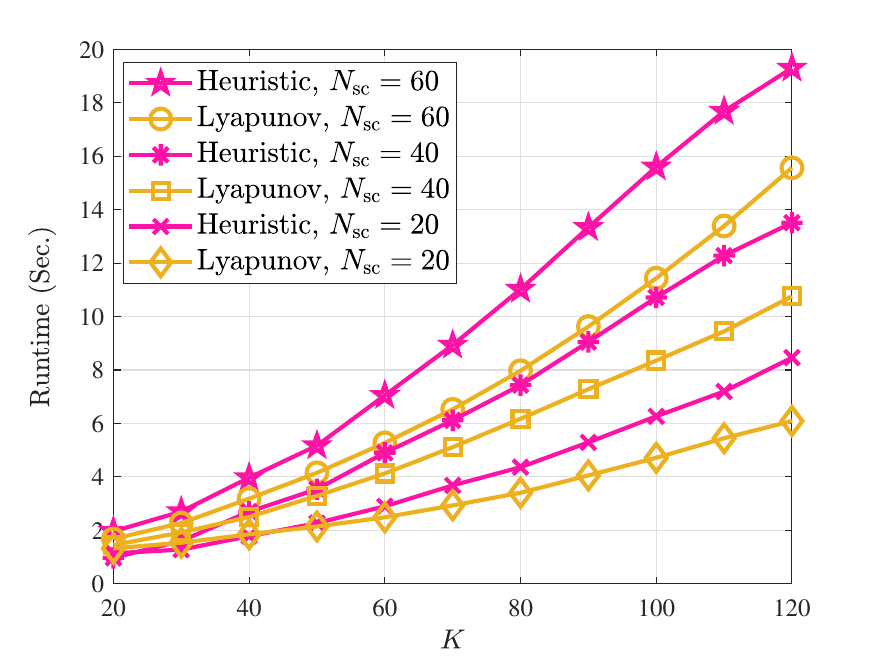}
\caption{Runtime of small-timescale RRA  methods.}
\label{fig:runtime}
\end{figure}

\textcolor{black}{Fig. \ref{fig:runtime} shows the runtime of our heuristic and the Lyapunov methods for small-timescale RRA executed by MATLAB 2024a on a computer with CPU of 2 GHz base clock and 16 GB RAM. The CC analysis in Subsection \ref{sec:cc_rra} shows that the RRA is relatively of cubic worst-case complexity with respect to $K$ and $N_{\rm sc}$. However, simulations show considerably lower complexity, as for example, six time increase in $K$ results in less than ten times increase in runtime in our heuristic methods. It is observable that the complexity is even lower with respect to $N_{\rm sc}$. In practice, the runtime may become orders of magnitude lower by using low-level and customized interior-point method implementations instead of high-level off-the-shelf solvers such as CVX.}

\section{\textcolor{black}{Conclusion \& Future Works}}\label{sec:conclusion}
\noindent In this paper, we proposed a mixed-timescale optimization framework for task offloading problem in vehicular networks. Specifically, we performed the CRA at the large-timescale with the aim to be robust against demand uncertainty, i.e., meeting the latency deadlines even for the largest possible tasks' data sizes. Besides, a closed-form expression of the optimal task partitioning ratio is obtained by noting that the latency constraints are active at the optimal solution point. The small-timescale RRA is performed by ensuring that each task receives the necessary data rate in each time slot to meet the latency constraint. The simulation results show the efficiency of the proposed method. Specifically, we showed that the conservativeness of the robust large-timescale CRA is perfectly compensated for by the small-timescale RRA. We also studied a trad-off between radio and computational resources, leading to a versatile system design that is capable of mitigating the resource shortage in one resource domain by allocating more resources in the other resource domain.

\textcolor{black}{Future research may include incorporating RL-based methods that reliably satisfy task deadlines. Although many RL-based methods are already studied in the context of task offloading, the mixed-timescale structure of resource allocation poses new challenges to existing RL approaches that need to be addressed.}
\ifCLASSOPTIONcaptionsoff
  \newpage
\fi



%
%
%

\bibliographystyle{ieeetr}
\balance
\bibliography{citation_v2x}{}

%
%
%
%

%
%
\begin{IEEEbiography}[{\includegraphics[width=1in,height=1.2in,clip,keepaspectratio]{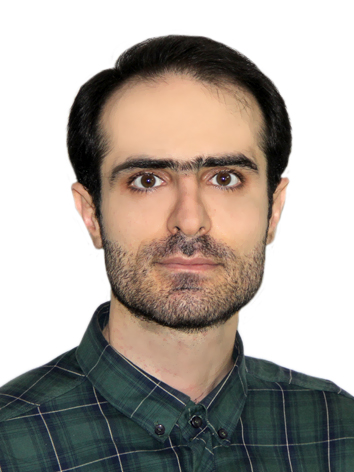}}]{Mohsen Tajallifar} received the B.Sc. degree in electrical engineering from Shiraz University, Shiraz, Iran, in 2010, and the M.Sc. and Ph.D. degrees in electrical engineering from Tarbiat Modares University, Tehran, Iran, in 2013 and 2022, respectively. He is currently a Postdoctoral Fellow with the Electrical Engineering Department, Qatar University, Doha, Qatar. He received the Best Paper Award at ITU Kaleidoscope 2020. His research interests include optimization methods, multiple-antenna communication systems, edge computing, and resource allocation in wireless networks.
\end{IEEEbiography}
\vspace{-1cm}
\begin{IEEEbiography}[{\includegraphics[width=1in,height=1.2in,clip,keepaspectratio]{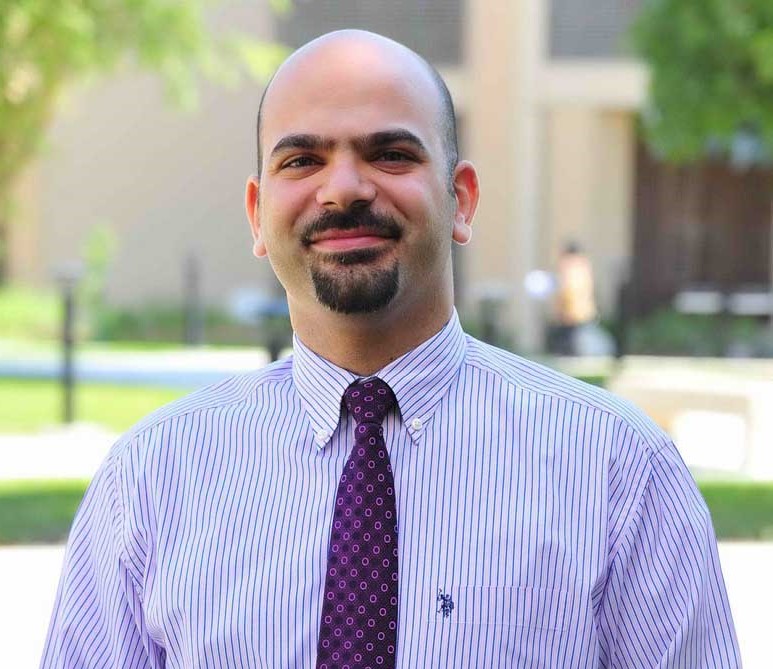}}]{Nizar Zorba} (Senior Member, IEEE) received the B.Sc. degree in electrical engineering from JUST University, Jordan, in 2002, and the Ph.D. degree in signal processing for communications from UPC Barcelona, Spain, in 2007. He is currently a Professor with the Electrical Engineering Department, Qatar University, Doha, Qatar. He has authored seven international patents, two books, contributed to seven book chapters, and co-authored over 200 papers in peer-reviewed journals and international conferences. He is associate EiC of the IEEE Communications Magazine, an Executive Editor of the IEEE Communications Letters and an Associate Editor of IEEE Transactions on Cognitive Communications and Networking. He served as an IEEE ComSoc Distinguished Lecturer 2024-2025. Dr. Zorba is a Former Chair of the IEEE Communication Society Technical Committee on Communications Systems Integration and Modelling.
\end{IEEEbiography}

\begin{IEEEbiography}[{\includegraphics[width=1in,height=1.2in,clip,keepaspectratio]{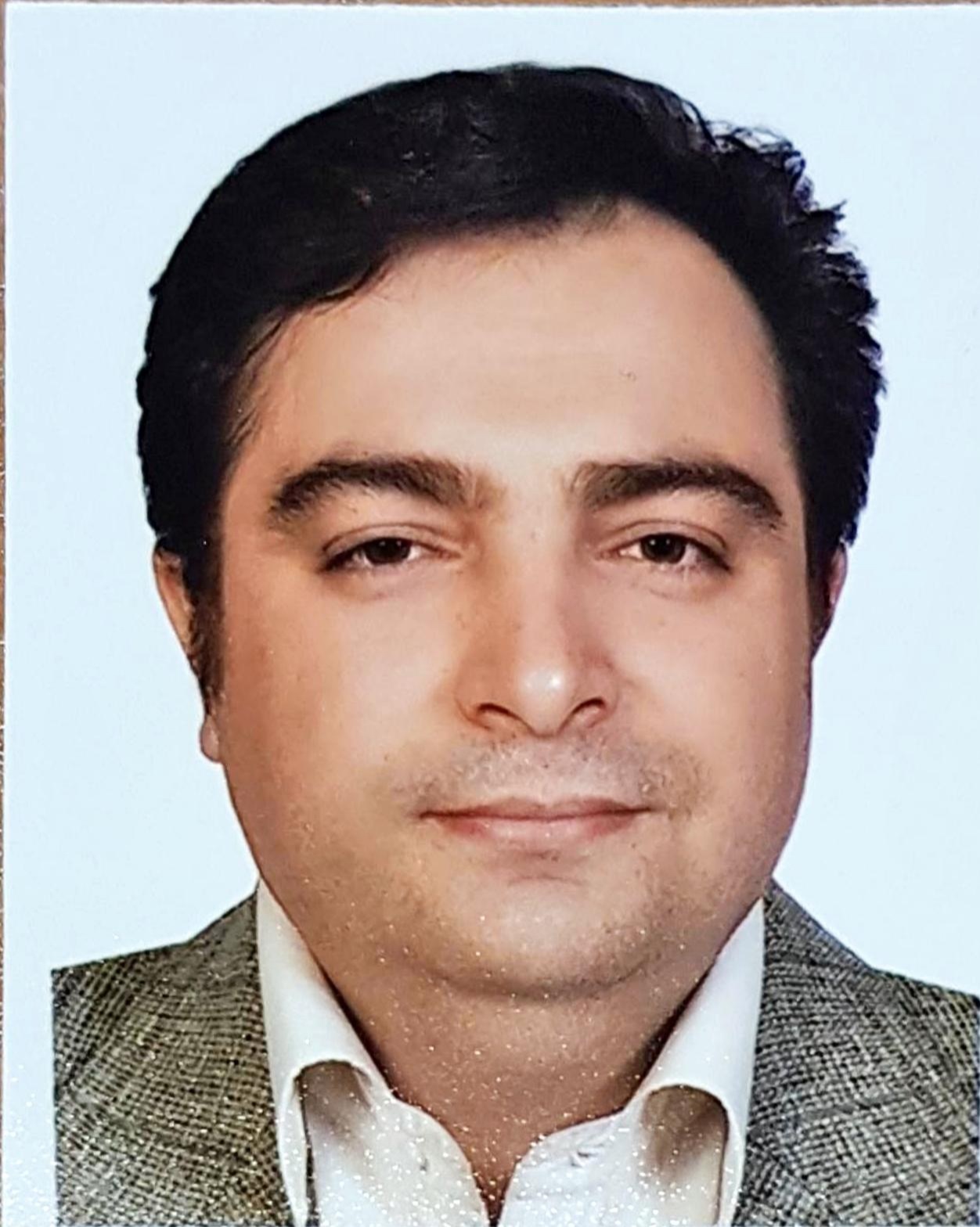}}]{Nader Mokari} (Senior Member, IEEE) received the Ph.D. degree in electrical engineering from Tarbiat Modares University, Tehran, Iran, in 2014. His thesis received the IEEE outstanding Ph.D. Thesis Award.  Since 2015, he has been with the Department of Electrical and Computer Engineering, Tarbiat Modares University, where he is currently an Associate Professor. His research interests cover many aspects of wireless technologies with a special emphasis on wireless networks. In recent years, his research has been funded by mobile telecommunication companies, the Iran National Science Foundation, German Research Foundation (DFG), and Qatar Research, Development and Innovation (QRDI) Council. He is also involved in a number of large scale network design and consulting projects in the telecom. industry. He is also the recipient of the Best Paper Award at ITU Kaleidoscope-2020. He has been elected as an IEEE Exemplary Reviewer in 2016 by IEEE Communications Society. He is currently serving as a member of the editorial board of the IEEE Transactions on Communications.
\end{IEEEbiography}

\begin{IEEEbiography}[{\includegraphics[width=1in,height=1.2in,clip,keepaspectratio]{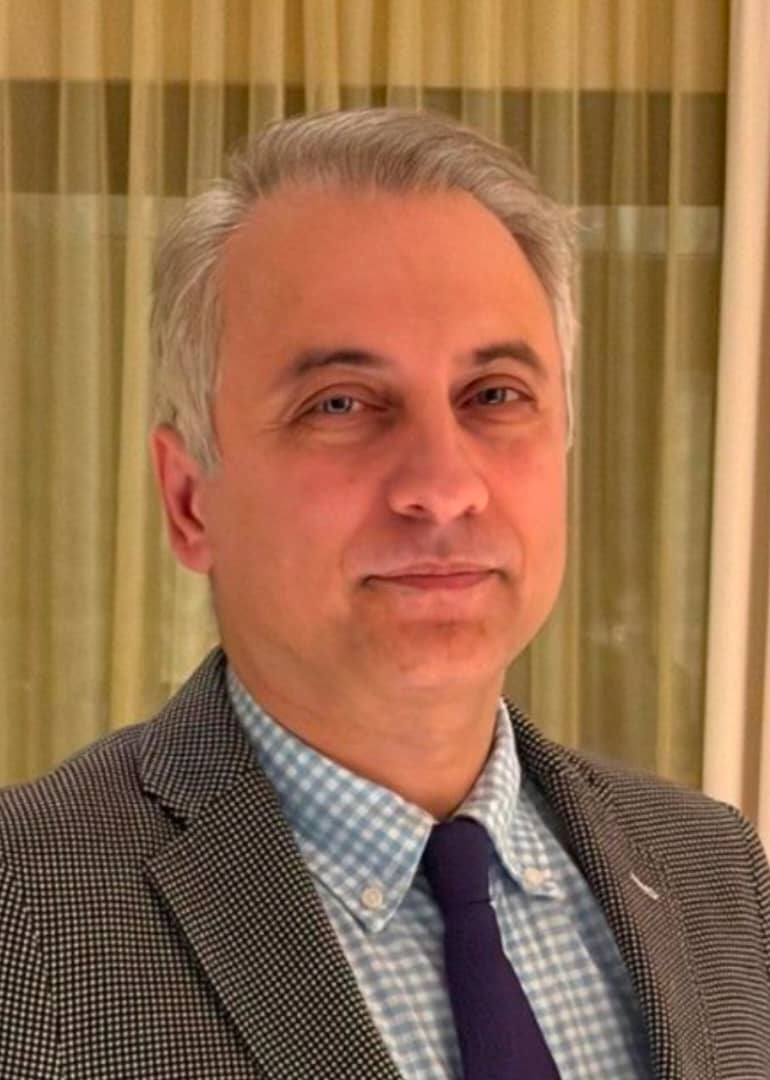}}]{Hamid Saeedi} (Member, IEEE) received the B.Sc. and M.Sc. degrees in electrical engineering from the Sharif University of Technology, Tehran, Iran, in 1999 and 2001, respectively, and the Ph.D. degree in electrical engineering from Carleton University, Ottawa, ON, Canada, in 2007. From 2008 to 2009, he was a Post-Doctoral Fellow with the Department of Electrical and Computer Engineering, University of Massachusetts at Amherst, Amherst, MA, USA. In 2010, he joined the Department of Electrical and Computer Engineering, Tarbiat Modares University, Tehran. Since 2021, he has been with the College of Engineering, University of Doha for Science and Technology, Doha, Qatar, as an Associate Professor. He was a member of the editorial board of IEEE Communications Letters and IEEE Transactions on Communications from 2014 to 2018 and from 2018 to 2020, respectively.
\end{IEEEbiography}

%
%
%




\end{document}